\documentclass[numbers,3p,11pt]{elsarticle}

\usepackage{amsmath,amssymb,mathtools}

\usepackage{ulem}
\usepackage{bm}

\newcommand{\myeqref}[2]{(\ref{#1}#2)}
\newcommand{\myref}[2]{\ref{#1}#2}
\newcommand{\subtag}[1]{\tag{\theparentequation #1}}

\newcommand*\de{\mathop{}\!\mathrm{d}}
\newcommand*{\Oh}{\mathcal{O}}
\newcommand*{\ie}{\emph{i.e.}}

\newcommand*{\etc}{\emph{etc}.}
\renewcommand*{\Re}{\operatorname{Re}}
\renewcommand*{\Im}{\operatorname{Im}}
\newcommand{\im}{\mathrm{i}}

\newcommand{\beq}{\begin{equation}}
\newcommand{\eeq}{\end{equation}}
\newcommand{\upd}{\mathrm{d}}
\newcommand{\kinf}{K_\infty}

\newcommand*{\dirac}{\delta_\mathrm{Dirac}}
\newcommand*{\kdill}{k_{\mathrm{I}}}
\newcommand*{\lD}{\ell_{\mathrm{I}}}
\newcommand*{\KD}{K_{\mathrm{I}}}
\newcommand*{\kwink}{k_{\mathrm{W}}}
\newcommand*{\lW}{\ell_{\mathrm{W}}}
\newcommand*{\phhat}{\hat{\phi}}
\newcommand*{\hhat}{\hat{h}}
\newcommand*{\Uhat}{\hat{U}}
\newcommand*{\ohat}{\hat{\omega}}
\newcommand*{\that}{\hat{t}}
\newcommand*{\Hcal}{\mathcal{H}}
\newcommand*{\C}{\mathcal{C}}
\newcommand*{\Chat}{\hat{\mathcal{C}}}
\newcommand*{\nabh}{\nabla_H}

\journal{Proceedings of the Royal Society}

\begin{document}

\begin{frontmatter}

\title{Validity of Winkler's mattress model for thin elastomeric layers: Beyond  Poisson's ratio}

%% use optional labels to link authors explicitly to addresses:
%% \author[label1,label2]{}
%% \address[label1]{}
%% \address[label2]{}

\author{%%%% Author details
Thomas G.~J.~Chandler and Dominic Vella}

\address{Mathematical Institute, University of Oxford, Woodstock Rd, Oxford, OX2 6GG, UK}

% \tableofcontents
% \newpage

\begin{abstract}
Winkler's mattress model is often used as a simplified model to understand how a thin elastic layer, such as a coating, deforms when subject to a distributed normal load: the deformation of the layer is assumed proportional to the applied normal load. This simplicity means that the Winkler model has found a wide range of applications from soft matter to geophysics. However, in the limit of an incompressible elastic layer  the model predicts infinite resistance to deformation, and hence breaks down. Since many of the thin layers used in applications are elastomeric, and hence close to incompressible, we consider the question of when the Winkler model is appropriate for such layers. We formally derive a model that interpolates between the Winkler and incompressible limits for thin elastic layers, and illustrate this model by detailed consideration of two example problems: the point indentation of a coated elastomeric layer and self-sustained lift in soft elastohydrodynamic lubrication. We find that the applicability (or otherwise) of the Winkler model is not determined by the value of the Poisson ratio alone, but by a compressibility parameter that combines the Poisson ratio with a measure of the layer's slenderness that depends on the problem under consideration.
\end{abstract}

%\keywords{elasticity, Winkler mattress model, perturbation analysis}

\end{frontmatter}

\section{Introduction}
%%%% Insert A head here

Thin elastic solids coating another, more rigid, substrate are encountered in a range of scenarios, and at a range of scales, from the consolidated till separating a glacier from bedrock \cite{Sayag2011,Butler2020} to multilayer microfluidic systems \cite{Thorsen2002}. In many of these scenarios, it is crucial to be able to determine how the coating deforms. The  simplest, and perhaps most common,  model of this type is the Winkler model \cite{Dillard2018}, in which the vertical displacement of the coating's interface, denoted $-\zeta(x)$, is proportional to the applied pressure, with some `Winkler modulus' $\kwink$,  \ie~we may write
\beq
-\zeta(x)=\frac{p(x)}{\kwink}.
\label{eq:Winkler}
\eeq 
On dimensional grounds, the stiffness $\kwink$ must be proportional to $G/d$ with $G$ the shear modulus of the layer and $d$ its thickness \cite{Johnson1985}. Using an analogue of hydrodynamic lubrication theory, Skotheim \& Mahadevan \cite{Skotheim2004} showed that, for a thin elastic solid, the prefactor is determined in terms of the Poisson ratio, $\nu$, by \beq
\kwink=\frac{2(1-\nu)}{1-2\nu}\frac{G}{d}.
\label{eqn:kwink}
\eeq 
Physically, the Winkler model corresponds to the elastic coating being made up of a series of disconnected springs, each of which deforms in proportion to the local pressure. For this reason, Winkler's model is often also referred to as the `mattress model'. While this is an extremely simple, as well as natural, model for many elastic deformations, the modulus $\kwink$ in \eqref{eqn:kwink} diverges as the Poisson  ratio $\nu\to1/2$:  an incompressible spring cannot deform with any finite applied pressure. Nevertheless, the Winkler model is often successfully used to describe experiments in which the thin coating is elastomeric and so might, ordinarily, be considered to be close to incompressible \cite{Saintyves2016,Rallabandi2017,Saintyves2020}. Why does the Winkler model work so well in such situations?

Of course, incompressible elastic materials \emph{are} able to deform in response to loading, even if this is contrary to the suggestion of the Winkler model. For an incompressible material,  deformation occurs via the shear of the material --- an effect that is missed by the Winkler model.  In fact, based on the work of Gent and Meinecke \cite{Gent1970}, Dillard \cite{Dillard1989} showed that the correct model for a one-dimensional, perfectly incompressible, coating is that the deformation of the interface is proportional to the second derivative of the pressure, \ie~that 
\beq
\zeta(x)=\frac{1}{\kdill}\frac{\upd^2p}{\upd x^2},
\label{eq:Dillard}
\eeq 
for some incompressible modulus $\kdill\propto G/d^3$ (where the coefficient of proportionality  depends on the precise boundary conditions assumed).

To generalize and combine these two limits, Lai \emph{et al.} \cite{Lai1992}  and Bert \cite{Bert1994} used an ansatz for the displacement field in which the horizontal displacement is both parabolic and up-down symmetric, while the vertical strain is assumed uniform. This assumption led to an expression for the vertical surface displacement, $\zeta$, that combines \eqref{eq:Winkler} and \eqref{eq:Dillard}, albeit with stiffness $\kwink$ that deviates from \eqref{eqn:kwink} for materials not close to incompressible. In this paper, we use asymptotic analysis, exploiting the slenderness of the coating, to combine the two results \eqref{eq:Winkler} and \eqref{eq:Dillard} more formally. We derive a combined foundation model that can be used when the response is not known, \emph{a priori}, to lie in either the Winkler or incompressible regimes and with no-slip or specified shear surface boundary conditions. In particular, we show that in this `combined foundation' model, the displacement in the normal, $z$, direction from the Winkler and incompressible models can be added to give $\zeta$ in terms of the surface pressure field $p(x,y)$ as
\beq
-\zeta_C=\frac{p}{\kwink}-\frac{1}{\kdill}\nabla_H^2p,
\label{eqn:CombFound}
\eeq 
where $\nabla_H^2=(\partial_x^2+\partial_y^2)$ is the two-dimensional Laplacian. We apply this combined foundation model to two examples and discuss the signatures of the transition between the Winkler and incompressible regimes, and how it might be identified experimentally.  We begin by considering the full three-dimensional formulation of the substrate deformation, before exploiting the coating's slenderness to derive \eqref{eqn:CombFound}.

\section{General formulation and analysis\label{sec:GenTheory}}

\subsection{Problem setup}

\begin{figure}
\centering
\includegraphics[width=.95\textwidth]{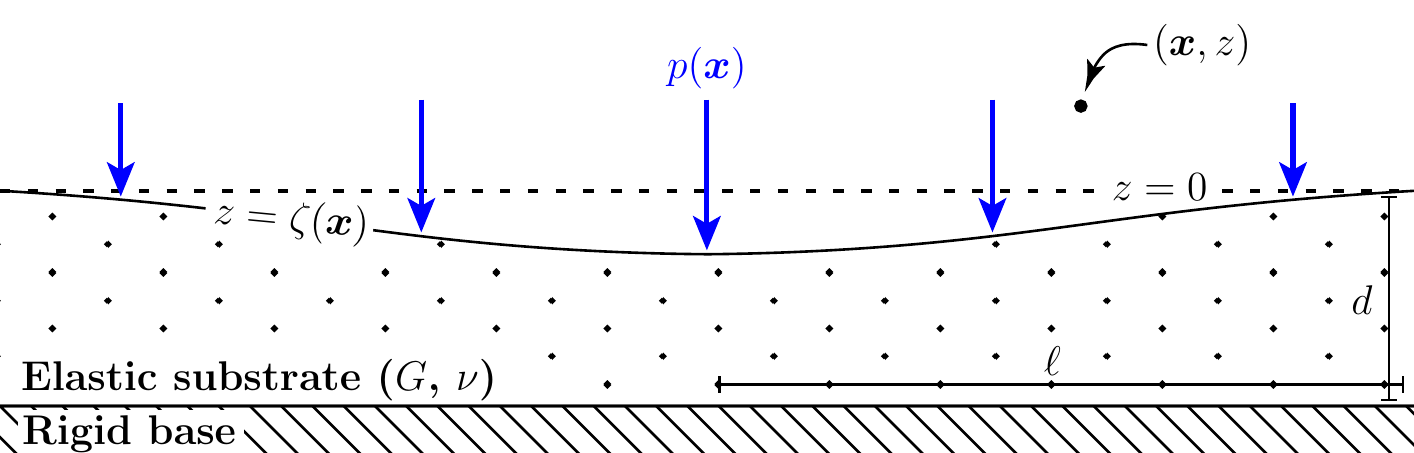}
\vspace*{10pt}
\caption{A cross-section sketch of the general setup considered in this paper: the surface (at $z=0$) of a three-dimensional elastic block adhered to a rigid base (at $z=-d$), with aspect ratio $\epsilon\coloneqq d/\ell\ll1$, is displaced by a pressure field $p(\bm{x})$. The position of the deformed surface  is  $z=\zeta(\bm{x})$.
\label{fig:setup}}
\end{figure}

Consider  a thin deformable layer (of thickness $d$, Poisson  ratio $\nu$, and shear modulus $G$) bonded to a rigid substrate. The surface of the deformable layer is subject to a given pressure field $p(\bm{x})$ and is displaced from $z=0$ to $z=\zeta(\bm{x})$; here, and throughout, $\bm{x}=(x,y)$ denotes Cartesian coordinates in the plane $z=0$. A cross-section of this setup is sketched in Fig.~\ref{fig:setup}. 

Small, steady displacements within the deformable layer may be modelled using steady linear elasticity: the planar and vertical displacements, $\bm{u}(\bm{x},z)=(u,v)$ and $w(\bm{x},z)$, respectively, are governed by the steady Navier--Cauchy equations \cite{Howell2008},
\begin{subequations}\label{eq:Navier_eq}
\begin{align}
\nabh (\nabh\cdot \bm{u}+ w_z) + (1-2\nu) \left(\nabh^2 \bm{u}+ \bm{u}_{zz}\right)&=\bm{0},\\
\frac{\partial }{\partial z} (\nabh\cdot \bm{u}+ w_z) + (1-2\nu) \left(\nabh^2 w+ w_{zz}\right)&=0.
\end{align}
\end{subequations}
Here (and henceforth) $\nabh=(\partial_x,\partial_y)$ denotes the planar del-operator and subscripts denote partial derivatives.

The deformable layer is assumed to be perfectly adhered to the base since this is the intention of most experiments. We therefore apply a zero-displacement boundary condition,
\begin{subequations}\label{eq:Navier_bcs}
\begin{equation}
\bm{u}=\bm{0} \quad \text{and} \quad w=0\quad \text{on} \quad z=-d. \subtag{a,b}
\end{equation}
The top surface, however, is assumed free, so we apply normal and tangential stress balances:
\begin{align}
\frac{2G}{1-2\nu}\left[ \nu\nabh \cdot\bm{u}+(1-\nu)w_z\right]\equiv\tau_{zz}&=-p(\bm{x}),\subtag{c}\\
G\left[ \bm{u}_z+\nabh w\right]\equiv(\tau_{xz},\tau_{yz})&=\bm{T}(\bm{x}),\subtag{d} \label{eq:Navier_bcs_shear}
\end{align}
\end{subequations}
on $z=0$, respectively. (Here we apply the free surface boundary conditions at $z=0$, consistent with the infinitesimal displacements of linear elasticity.) 

In \eqref{eq:Navier_bcs_shear}, $\bm{T}(\bm{x})$ denotes the two-dimensional vector of  shear stresses applied to the surface. Often,  $\bm{T}(\bm{x})$ may be coupled to the normal loading, namely the pressure $p(\bm{x})$, while in other circumstances, a particular shear stress distribution is required to ensure a particular boundary condition on the surface $z=0$ is satisfied. We shall consider specific examples of this coupling in the two  problems considered in \S\ref{sec:FirstExample} and \S\ref{sec:SecondExample}.  For now, however, we consider the problem for general pressure  and shear stress profiles, $p(\bm{x})$ and $\bm{T}(\bm{x})$. 

Solving \eqref{eq:Navier_eq} subject to the boundary conditions \eqref{eq:Navier_bcs} is difficult and cannot, in general,  be done analytically; progress can be made, however,  in the case of a two-dimensional substrate or a rotationally-symmetric substrate by using Fourier or Hankel transforms, respectively \cite{sneddon1995}. In many applications of interest, however,  the deformable  layer has a small aspect ratio $\epsilon\coloneqq d/\ell\ll1$ (where $\ell$ is the horizontal length-scale over which the substrate is deformed) \cite{Chen2005, Hyun2009,Kaltenbrunner2013,Davies2018, Saintyves2016, Saintyves2020, Zhang2020}. We therefore turn now to consider how this small aspect ratio can facilitate analytical progress.

\subsection{Thin layer analysis\label{sec:ThinAnalysis}}

The observation that many applications have a small aspect ratio, together with the success of lubrication theory in viscous fluid mechanics \cite{Leal2007}, motivates the posing of an asymptotic expansion as $\epsilon\to 0$. In \ref{app:asymp_deriv} we present the formal asymptotic expansion   under the assumption of a thin geometry: $d=[z]=\epsilon[\bm{x}]$. In this way, we obtain an asymptotic series for the displacements $\bm{u}(\bm{x},z)$ and $w(\bm{x},z)$ in powers of  $\epsilon$ that depends on the pressure and shear profiles, $p(\bm{x})$ and $T(\bm{x})$, respectively. In dimensional terms, the  displacements of the surface are given by:
\begin{subequations}\label{eq:Chandler_full}
\begin{align}
\bm{u}(\bm{x},0)&= \frac{d}{G}\bm{T}(\bm{x}) - \beta(\nu)\frac{ d^2}{G}\nabh p(\bm{x}) + \Oh(\ell\epsilon^3),\\
w(\bm{x},0)&= -\alpha(\nu)(1-2\nu)\frac{d}{G}p(\bm{x}) -\beta(\nu)\frac{d^2}{G}\nabh\cdot \bm{T}(\bm{x})+ \gamma(\nu)\frac{d^3}{G}\nabh^2p(\bm{x}) + \Oh(d\epsilon^3),
\end{align}
\end{subequations}
where, for convenience, we introduce the coefficients
\begin{subequations}\label{eq:abg_def}
\begin{equation}
\alpha(\nu)\coloneqq \frac{1}{2(1-\nu)},\qquad
\beta(\nu)\coloneqq \frac{\nu-1/4}{1-\nu},\qquad
\gamma(\nu)\coloneqq \frac{2\nu(\nu-1/4)}{3(1-\nu)^2}. \subtag{a--c}
\end{equation} 
\end{subequations}

Equation \myeqref{eq:Chandler_full}{b} is the main result of this paper: the surface deformation of a  thin deformable layer is a linear combination of the applied normal pressure $p(\bm{x})$, the tangential shear stress $\bm{T}(\bm{x})$, and their spatial derivatives.  We note here that \eqref{eq:Chandler_full} may not hold close to boundaries if the imposed boundary conditions are incompatible with the displacement field suggested by \eqref{eq:Chandler_full}; in such a scenario, we expect a  boundary layer of size $[\bm{x}]\sim d$ to resolve this incompatibility. The examples we consider in this paper avoid this incompatibility, and so analysis of such boundary layers is left for future work.

The quantity of most interest in applications is the vertical displacement of the surface, $\zeta(\bm{x})\coloneqq w(\bm{x},0)$. While this can be determined very simply from \eqref{eq:Chandler_full}, we consider two common cases in some detail: no (or negligible) surface-shear, $\bm{T}(\bm{x})\equiv \bm{0}$, and zero surface-slip, $\bm{u}(\bm{x},0)\equiv\bm{0}$. In these cases the deformed surface profile \myeqref{eq:Chandler_full}{b} simplifies to the expression,
\begin{equation}\label{eq:Chandler_found}
\zeta(\bm{x})\coloneqq w(\bm{x},0)\sim -\alpha(\nu)(1-2\nu)\frac{d}{G}p(\bm{x})+ \Gamma(\nu)\frac{d^3}{G}\nabh^2p(\bm{x}),
\end{equation}
where
\begin{equation}\label{eq:fuctions_fg}
\Gamma(\nu)\coloneqq\begin{dcases}
\gamma(\nu)=\frac{2\nu(\nu-1/4)}{3(1-\nu)^2} &\text{for zero surface-shear,}\\
\gamma(\nu)-\beta(\nu)^2=\frac{(3/4-\nu)(\nu-1/4)}{3(1-\nu)^2} &\text{for zero surface-slip.}
\end{dcases}
\end{equation}
We note that for incompressible substrates, the prefactor $\Gamma(\nu)$ for zero surface-shear is precisely 4 times that for zero surface-slip, as is familiar in hydrodynamic lubrication theory \cite{Leal2007, Mukherjee2018}. However, for  compressible substrates, $\nu\neq 1/2$, the ratio of these factors is $\nu$-dependent, because the  displacement and stress profiles are not symmetric about $z=-d/2$ , see \ref{app:asymp_deriv}, especially \eqref{eq:U0W0_sol}~\&~\eqref{eq:W1_sol}.  Note also that although the incompressible response of a thin elastomer, \eqref{eq:Dillard}, can be directly compared to the classical results of lubrication theory for an incompressible fluid \cite{Gent1970}, specifically Reynolds' equation \cite{Leal2007}, we are not aware of an  analogue of \eqref{eq:Winkler} for a  compressible fluid with spatially varying pressure \cite{Almqvist2019}.

The surface deflection described in \eqref{eq:Chandler_found} is precisely a linear superposition of the Winkler and incompressible foundation models \eqref{eq:Winkler} and \eqref{eq:Dillard}. While this result might seem intuitive, here we have derived it formally.  Our formal asymptotic technique is more rigorous than previous work~\cite{Lai1992, Bert1994}, in which a parabolic displacement field and uniformly distributed out-of-plane strain were posed as an ansatz to solve the problem. Moreover, our technique allows us, in general, to consider the effect of surface shear forces, and different boundary conditions on the surface of the substrate. Nonetheless, the relevant results of ref.~\cite[eq.~(24)]{Lai1992} and  ref.~\cite[eq.~(10)]{Bert1994} are recovered in the incompressible limit ($\nu\to 1/2$) of \eqref{eq:Chandler_found}, for the zero surface-slip case.

To understand the relative importance of the Winkler and incompressible responses, we estimate the ratio between the respective terms in \eqref{eq:Chandler_found} as $(1-2\nu)/\epsilon^2$: for sufficiently compressible materials, $ (1-2\nu) \gg \epsilon^2 $, the surface deflection is dominated by the $p(\bm{x})$-term and the layer is well-described by the  Winkler foundation model. However, for materials that are `close' to incompressible, $ (1-2\nu)\ll \epsilon^2 \ll 1$, the substrate response is dominated by the $\nabh^2 p(\bm{x})$-term, and the layer instead behaves according to   the incompressible elastomeric foundation model. 

Our combined model of the layer deflection shows that whether the layer behaves as an incompressible or a compressible layer depends not simply on the value of its Poisson ratio $\nu$, but also on its slenderness: even layers that should be expected to be incompressible on the basis of their Poisson ratio, \ie~$\nu\approx1/2$, may behave in a compressible manner if they are sufficiently thin that  $\epsilon^2\lesssim(1-2\nu)$. Since quoted values of $\nu$ for elastomeric layers suggest $0.45\lesssim\nu\lesssim0.5$\footnote{Given this range of $\nu$, we do not consider the possibility that $\Gamma(\nu)=0$, which occurs when $\nu=1/4$.} \cite{Dogru2018}, we typically expect $1-2\nu\sim10^{-1}$ and hence it is clearly feasible that in applications elastic layers will be sufficiently thin to behave in a compressible manner, \ie~the Winkler model may still be appropriate.  Further, one can readily imagine an experiment in which both  compressible (Winkler's) and incompressible  responses are observed with slightly different parameter values.

Of course, the above discussion depends on the value of the parameter $\epsilon$, which in turn depends on the length scale $\ell$. The challenge, however, is that $\ell$ is not, in general, known \emph{a priori}; instead, $\ell$ must be determined as part of the solution of a given problem. In the remainder of this paper, we illustrate this dependence, and discuss when Winkler, rather than incompressible behaviour, can be expected in two concrete examples: the point-like indentation of a thin elastic layer with a stiff coating  (see \S\ref{sec:FirstExample}) and the elastohydrodynamic lift generated by a thin liquid layer above a thin elastic substrate (see \S\ref{sec:SecondExample}).

\section{First example: A soft layer with a stiff coating}\label{sec:FirstExample}

As a first example of the application of the combined foundation model, we consider the response of a relatively stiff elastic layer coating a much softer layer (the elastomer). Such coatings are used as a way to reinforce soft materials \cite{Takei2011} or to combine electrical conductivity with material flexibility, as in flexible electronics \cite{Rogers2009,Kaltenbrunner2013}. Motivated by the increase in effective stiffness of the substrate created by using a stiff coating, we consider the response to point-indentation. This is a problem that has been considered recently \cite{Cao2018, Niu2018,Liu2019, Box2020} for small indentations of a coated  elastic half-space (\ie~a substrate of infinite depth); in this limit it was shown \cite{Box2020} that the applied point-force $F$ required to produce a localized indentation depth $\delta$ is $F=\kinf\delta$, where the constant indentation stiffness
\beq
\kinf=\frac{3^{3/2}}{(1-\nu)^{2/3}}B^{1/3}G^{2/3},
\label{eqn:BoxStiffness}
\eeq
with $B$ the bending stiffness of the coating and $G$ and $\nu$ denoting the moduli of the substrate.

Hertz \cite{Hertz1884} considered the point-indentation of an elastic plate with a substrate response that is linear in vertical displacement --- while this response was envisaged to arise from the hydrostatic pressure in  liquid underlying floating ice, it is mathematically identical to the indentation of a Winkler foundation coated by a plate, and hence represents the thin, compressible analogue of \eqref{eqn:BoxStiffness}. In our notation, Hertz showed that the indentation stiffness $K=F/\delta=8(B\kwink)^{1/2}$, where $\kwink$ is the Winkler modulus of the substrate, as defined in \eqref{eq:Winkler}. In this section, we study how these results are modified when the substrate is slender and may be close to incompressible. 

\subsection{Mathematical modelling}

We consider small, axisymmetric deformations of an infinite plate (bending stiffness $B$) adhered to a thin substrate (of depth $d$, Poisson  ratio $\nu$, and  shear modulus $G$) in response to  point-like indentation. Note that  the coating could alternatively be parameterized by its thickness $t$, Poisson ratio $\nu_c$, and shear modulus $G_c$, by replacing
\begin{equation}\label{eq:stiffcoat_B}
B\equiv \frac{G_c t^3}{6(1-\nu_c)}.
\end{equation}

\begin{figure}[ht!]
\centering
\includegraphics[width=.8\textwidth]{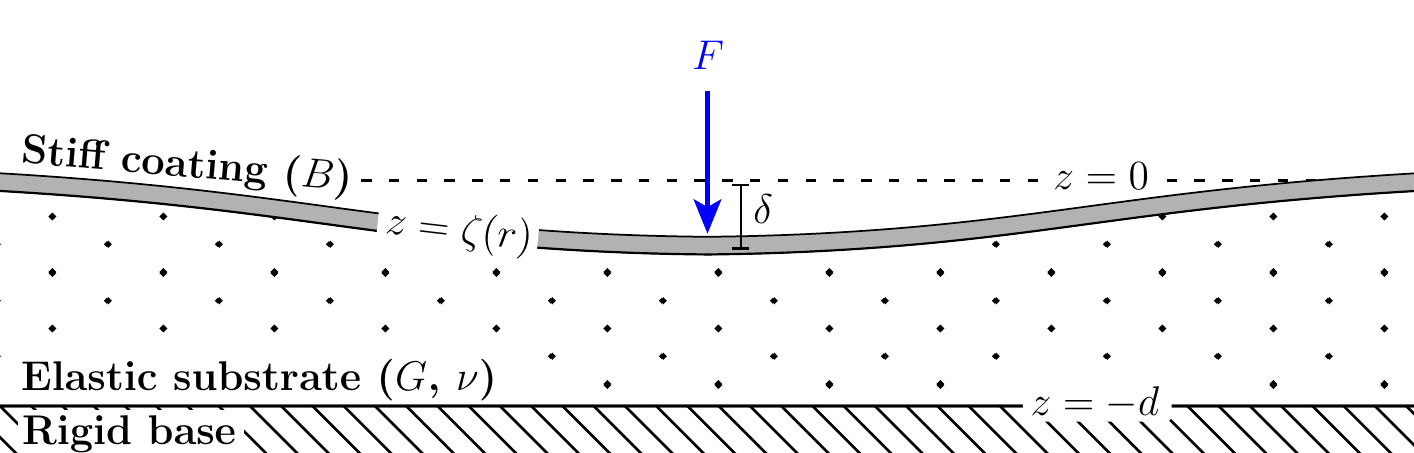}
\caption{A sketch of  the indentation of a stiff coating of a deformable substrate. The coating is loaded  by a point-force, $F$, applied at  $r=0$, resulting in a vertical displacement $\delta$; the vertical deflection  profile $\zeta(r)$ decays away from the indenter.
\label{fig:IndSetup}}
\end{figure}

The setup we consider is sketched in Fig.~\ref{fig:IndSetup}. Vertical displacements of the plate, $\zeta(r)$, in response to indentation and the restoring pressure, $q(r)$, from the underlying substrate are coupled using the zero-stretching Kirchhoff--Love plate equation \cite[Chap.~4]{Howell2008} together with the combined foundation model [\ie~taking $p(\bm{x})\mapsto q(r)$ in \eqref{eq:Chandler_found}]. We, therefore, have that
\begin{subequations}\label{eq:stiffcoat_eqs}
\begin{align}
B\nabh^4 \zeta(r) &= q(r)-\frac{F}{2\pi}\frac{\dirac(r)}{r}   ,\label{eq:stiffcoat_eqs_plate}\\
\zeta(r) &= -\alpha(\nu)(1-2\nu)\frac{d}{G} q(r)+ \Gamma(\nu)\frac{d^3}{G} \nabh^2 q(r),\label{eq:stiffcoat_eqs_chand}
\end{align}
\end{subequations} 
with $r>0$ the radial coordinate and $F$ the applied force. 

The plate is assumed to be perfectly adhered to the substrate and, since vertical deflections are small, we neglect horizontal displacements of the plate, which are no larger than quadratic in $\delta$ \cite{Vella2018,Box2020}, so that the boundary condition on the upper surface of the substrate is $u(r,0)\equiv 0$; we therefore use $\Gamma(\nu)$ to denote the coefficient given in \eqref{eq:fuctions_fg} for zero surface-slip. We seek solutions of \eqref{eq:stiffcoat_eqs} subject to the boundary conditions:
\begin{subequations} \label{eq:stiffcoat_bcs}
\begin{gather}
\zeta(0)=-\delta,\quad \zeta'(0)=0, \quad  q'(0)= 0,\subtag{a--c}\\
 \zeta(r),\, q(r) \to  0 \qquad \text{as $r\to \infty$}.\subtag{d,e}
\end{gather}
\end{subequations}
Equation \myeqref{eq:stiffcoat_bcs}{a} enforces the indentation depth $\delta$ while \myeqref{eq:stiffcoat_bcs}{b,c} ensure symmetry about the indentation point; the conditions \myeqref{eq:stiffcoat_bcs}{d,e} ensure decay far from the indenter.

\subsubsection{Scaling laws}

As a first step, we seek to determine the scaling behaviour of the radial scale, $\ell$. For a given indentation depth $\delta$, the length scale $\ell$ is chosen to minimize the sum of two opposing elastic energies \cite{Box2020}: the bending energy of the plate,  $U_\mathrm{plate}\sim B\int(\delta/\ell^2)^2 \de A \sim B(\delta/\ell)^2$, penalizes the curvature of the interface $ \sim \delta/\ell^2$ over the deformed area $A\sim \ell^2$ and so favours large $\ell$. In contrast,  the substrate's elastic energy, $U_\mathrm{subs}\sim \int \mathcal{T}\mathcal{E} \de V\sim \mathcal{T}\mathcal{E}(\ell^2  d) $, penalizes the  strain energy density $\sim \mathcal{T}\mathcal{E}$ within the deformed volume $V\sim \ell^2d$ and, as we shall see, favours small $\ell$. There is, therefore, an optimal  $\ell$ that minimizes the total elastic energy $U_\mathrm{elast}=U_\mathrm{plate}+U_\mathrm{subs}$. However, the result of this minimization depends on the typical size of the stress and strain, $\mathcal{T}$  and $\mathcal{E}$, within the substrate, which are in turn  dependent on the deformation mode of the substrate (\ie~how compressible it is). For sufficiently compressible substrates (the Winkler response), the deformation is dominated by the vertical bulk-strain, so that $\mathcal{T}\sim G\mathcal{E}_{zz}/(1-2\nu)$ with $\mathcal{E}\sim\mathcal{E}_{zz}\sim \delta/d$, leading  to $U_\mathrm{subs}\sim  G/(1-2\nu) \ell^2 \delta^2/d $. For sufficiently incompressible substrates (incompressible response), the deformation is instead dominated by the shear-strain associated with a horizontal displacement $u$, which may be estimated as $u\sim \delta \ell/d$  from incompressibility. We therefore have
$\mathcal{T}\sim G\mathcal{E}_{rz}$ with $\mathcal{E}\sim\mathcal{E}_{rz}\sim u/d \sim \delta\ell/d^2$ and, hence, $U_\mathrm{subs}\sim G \ell^4 \delta^2/d^3  $. Minimizing the total elastic energy in each of the two cases gives: 
\begin{equation}
\ell \propto 
\begin{dcases}
 d\left[\frac{B}{G d^3}(1-2\nu)\right]^{1/4}\qquad&\text{(Winkler)},\\
 d\left[\frac{B}{G d^3}\right]^{1/6} \qquad&\text{(incompressible)}.
\end{dcases}\label{eq:example1_ellscale}
\end{equation}
(Note from \eqref{eq:stiffcoat_B} that 
\begin{equation}
\frac{B}{Gd^3} \equiv \frac{1}{6(1-\nu_c)}\frac{G_c}{G}\left(\frac{t}{d}\right)^3,
\end{equation}
so that $B/Gd^3$ combines the ratio of mechanical properties, $G_c/G$, with the geometrical ratio $t/d$.) Following this balance, the total elastic energy in each case  scales according to $U_\mathrm{elast}\sim B\delta^2/\ell^2$ --- the work done to reach this state must be done by the indenter so that $U_\mathrm{ind}\sim F\delta\sim U_\mathrm{elast}$ \cite{Box2020}, which therefore gives a constant  stiffness response  with:
\begin{equation}
\frac{F}{\delta}\sim \frac{B}{\ell^2}\propto 
\begin{dcases} 
 \left[\frac{G B }{d (1-2\nu)}\right]^{1/2}\qquad&\text{(Winkler)},\\
 \left[\frac{G B^2}{ d^3 }\right]^{1/3} \qquad & \text{(incompressible)}.
\end{dcases}\label{eq:example1_stiffscale}
\end{equation}
These results give a first understanding for how the elastic body responds in both the Winkler and incompressible regimes. However, our aim in this paper is to understand the transition between the two regimes beyond a scaling analysis; we therefore  turn now to this, beginning by non-dimensionalizing the model problem \eqref{eq:stiffcoat_eqs}--\eqref{eq:stiffcoat_bcs}.

\subsubsection{Non-dimensionalization}

Although the natural choice for dimensionless pressure and surface displacement follows from the plate equation \eqref{eq:stiffcoat_eqs_plate} and \myeqref{eq:stiffcoat_bcs}{a}, the choice of radial scale depends on the substrate response, as in \eqref{eq:example1_ellscale}. From the substrate equation \eqref{eq:stiffcoat_eqs_chand}, we hence observe that there are two natural choices:
\begin{subequations}\label{eq:stiffcoat_LDLW}
\begin{equation}
\lW \coloneqq d\left[\frac{B}{G d^3}\alpha(\nu)(1-2\nu)\right]^{1/4}
\quad \text{and} \quad
\lD \coloneqq d\left[\frac{B }{G d^3}\Gamma(\nu)\right]^{1/6}.
\subtag{a,b}
\end{equation}
\end{subequations}
We refer to the lengths $\lW$ and $\lD$ as the Winkler and incompressible lengths, respectively.

For definiteness, and without loss of generality, we choose the incompressible  length as our radial scale, \ie~we let $\ell\equiv [r]=\lD$, and hence define dimensionless variables
\begin{subequations}\label{eq:stiffcoat_dimvars}
\begin{equation}
\rho \coloneqq  \frac{r}{\lD}, \qquad Z(\rho)\coloneqq \frac{\zeta(r)}{\delta}, \qquad Q(\rho) \coloneqq \frac{q(r)}{B\delta/\lD^4}.
\subtag{a--c}
\end{equation}
\end{subequations}
With the non-dimensionalization \eqref{eq:stiffcoat_dimvars} the system, \eqref{eq:stiffcoat_eqs}--\eqref{eq:stiffcoat_bcs}, becomes
\begin{subequations}\label{eq:stiffcoat_eqs_nondim}
\begin{align}
\nabh^4 Z(\rho)&=Q(\rho) -\KD\frac{\dirac(\rho)}{\rho},\\
Z(\rho) &= -\C Q(\rho)+ \nabh^2 Q(\rho),
\end{align}
\end{subequations}
subject to
\begin{subequations}\label{eq:stiffcoat_bcs_nondim}
\begin{gather}
Z(0)=-1,\quad Z'(0)=0, \quad  Q'(0)= 0,\subtag{a--c}\\
Z(\rho),\; Q(\rho) \to  0 \qquad  \text{as $\rho\to \infty$}.\subtag{d,e}
\end{gather}
\end{subequations}
Note that this non-dimensionalization introduces two key dimensionless parameters
\begin{subequations}
\begin{align}
\C&\coloneqq \left(\frac{\lW}{\lD}\right)^4 =(1-2\nu)\frac{\alpha(\nu)}{\Gamma(\nu)^{2/3}} \left(\frac{B}{G d^3}\right)^{1/3},\label{eq:stiffcoat_N}\\
\mathrm{and}\quad\KD&\coloneqq\frac{F \lD^2}{2\pi B\delta}=\frac{\Gamma(\nu)^{1/3}}{2\pi }\left(\frac{d^3 }{GB^2}\right)^{1/3}\times \frac{F}{\delta}.\label{eq:stiffcoat_KD}
\end{align}
\end{subequations}
Here, $\C\propto (1-2\nu)$ measures how far from perfect incompressibility the substrate is --- it is a measure of the \emph{substrate compressibility}.

The second dimensionless parameter, $\KD$, is a dimensionless indentation stiffness. Since the problem is linear, only a particular value of $\KD$ will allow the boundary condition \myeqref{eq:stiffcoat_bcs_nondim}{a} to be satisfied for a given value of $\C$;  determining $\KD(\C)$ is the primary aim of our analysis.  However,  the general behaviour of $\KD$ as the substrate compressibility, $\C$, varies can be outlined: for $\C\ll1$,  the substrate is effectively incompressible and $\lD$ is the appropriate horizontal length scale, so we  expect $\KD\sim \text{const}$. For $\C\gg1$, the substrate is highly compressible and so $\lW$ is expected to be the appropriate horizontal length scale --- to cancel the factor $\lD^2$ in the definition of $\KD$,  we expect  to find $\KD\propto \C^{-1/2}$. These expectations are both consistent with the scaling results \eqref{eq:example1_stiffscale}; now we turn to analytical solutions of \eqref{eq:stiffcoat_eqs_nondim} subject to \eqref{eq:stiffcoat_bcs_nondim} to determine the relevant prefactors.

\subsection{Analytical Results}

\subsubsection{General solution}

The axisymmetric problem \eqref{eq:stiffcoat_eqs_nondim} may be solved directly using Hankel transforms \cite{sneddon1995}. An alternative approach \cite{Vella2012,Box2017} is to seek solutions in terms of modified Bessel functions of the second kind --- \ie~to let $Z(\rho) \propto K_0[\lambda \rho]$ with $\Re{\lambda}>0$ chosen to ensure that the far-field condition, $Z(\rho\to\infty)=0$, is satisfied. Since $\nabh^2Z=\lambda^2Z$ by assumption, we obtain an auxiliary equation:
\begin{equation}\label{eq:plate_auxiliary}
(\lambda^2)^3-\C (\lambda^2)^2-1=0.
\end{equation} 
An analysis of this equation shows that for $\C>0$ (\ie~$\nu<1/2$) there is only one real solution for $\lambda^2$ and so we may write the general solution as
\begin{subequations}
\begin{align}
Z(\rho) &= c_1 K_0[\lambda \rho]+ c_2 \Re\left\{K_0[\Lambda \rho]\right\}+c_3 \Im\left\{K_0[\Lambda \rho]\right\},\\
Q(\rho)&=\nabh^4 Z(\rho) =  c_1 \lambda^4 K_0[\lambda \rho] + c_2 \Re\left\{\Lambda^4 K_0[\Lambda \rho]\right\} + c_3 \Im\left\{\Lambda^4 K_0[\Lambda\rho]\right\},
\end{align}
for unknown real constants $c_i$ and roots of \eqref{eq:plate_auxiliary}, $\lambda$ and $\Lambda$,  chosen such that $\arg{\lambda}=0$ and $0<\arg{\Lambda}\leq\pi/2$. Further, \eqref{eq:plate_auxiliary} may be used to write the complex root $\Lambda$ in terms of  $\lambda$:
\begin{equation}
\Lambda^2 =\frac{\im\sqrt{4\lambda^6-1}-1}{2\lambda^4}, 
\end{equation} 
where $\lambda^2$ is determined by solving the cubic, \eqref{eq:plate_auxiliary}, and may be written explicitly as
\begin{align}
\lambda^2=\frac{1}{3}\left(\C+\frac{\C^2}{\Delta(\C)}+\Delta(\C)\right) \quad \text{with} \quad
\Delta(\C) \coloneqq \left[\C^3 +\frac{3}{2}\left(9+\sqrt{81 +12 \C^3}\right) \right]^{1/3}.\subtag{d,e}
\end{align}
\end{subequations}

The constants, $c_i$, are determined by imposing the boundary conditions  \myeqref{eq:stiffcoat_bcs_nondim}{a--c} at $\rho=0$; we find:
\begin{subequations}\label{eq:stiffcoat_fullsol}
\begin{equation}
Z(\rho) = \dfrac{\Im\left\{\frac{K_0[\lambda \rho]-K_0[\Lambda \rho]}{\lambda^4-\Lambda^4}\right\}}{\Im\left\{\frac{\log(\lambda/\Lambda)}{\lambda^4-\Lambda^4}\right\}} 
\qquad \text{and}\qquad 
Q(\rho) = \dfrac{\Im\left\{\frac{\lambda^4 K_0[\lambda \rho]-\Lambda^4 K_0[\Lambda \rho]}{\lambda^4-\Lambda^4}\right\}}{\Im\left\{\frac{\log(\lambda/\Lambda)}{\lambda^4-\Lambda^4}\right\}}.
\subtag{a,b}
\end{equation} 
 The indentation stiffness $\KD$ is calculated from the resultant shear force at the origin  to be
\begin{equation}\label{eq:stiffcoat_fullsol_stiff}
\KD = \dfrac{\Im\left\{\frac{1}{\lambda^2+\Lambda^2}\right\}}{\Im\left\{\frac{\log(\lambda/\Lambda)}{\lambda^4-\Lambda^4}\right\}}.\subtag{c}
\end{equation}
\end{subequations} 

\begin{figure}[ht!]
\centering
\includegraphics[width=0.8\textwidth]{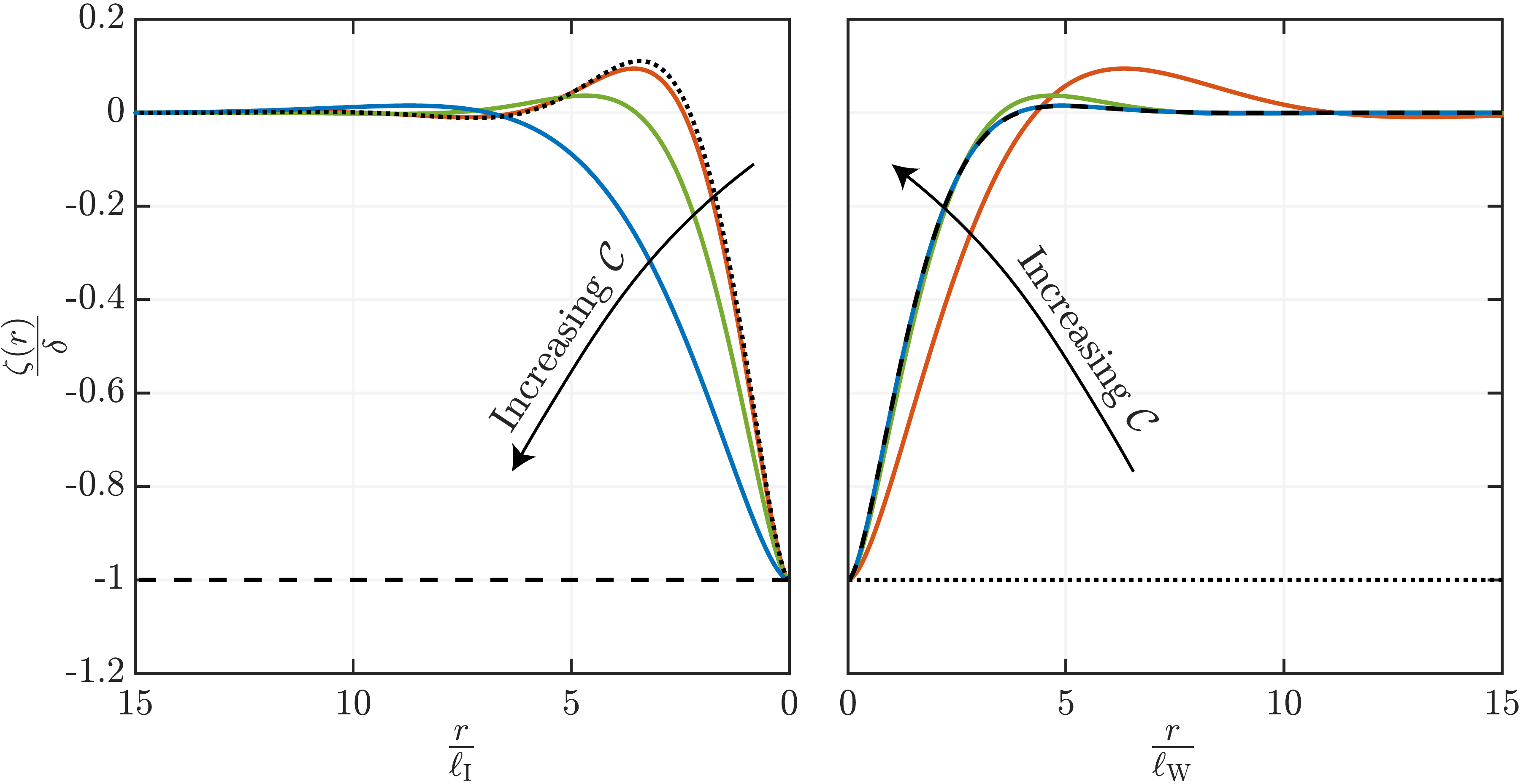}
\caption{The surface deflection of a stiff coating of a thin deformable substrate, as predicted by the combined foundation model for different values of the compressibility parameter $\C$ defined in \eqref{eq:stiffcoat_N}. Left: deflection profiles rescaled by the incompressible length scale, $\lD=(\Gamma(\nu)Bd^3/G )^{1/6}$ (as appropriate for the perfectly incompressible case). Right: deflection profiles rescaled by the Winkler length scale, $\lW=(\alpha(\nu)(1-2\nu)Bd/G )^{1/4}$ (as  appropriate for the compressible case). In each case, dashed curves show the interface profile predicted for a Winkler foundation, \eqref{eq:IndWinkProf}, while dotted curves show the interface profile predicted for an incompressible foundation by \eqref{eq:IndDillProf}. Solid curves show the  profiles predicted by the combined foundation model \eqref{eqn:CombFound}  for  $\C=0.1$ (red), $\C=1$ (green) and $\C=10$ (blue).
\label{fig:IndentationShapes}}
\end{figure}

\begin{figure}[ht!]
\centering
\includegraphics[width=0.7\textwidth]{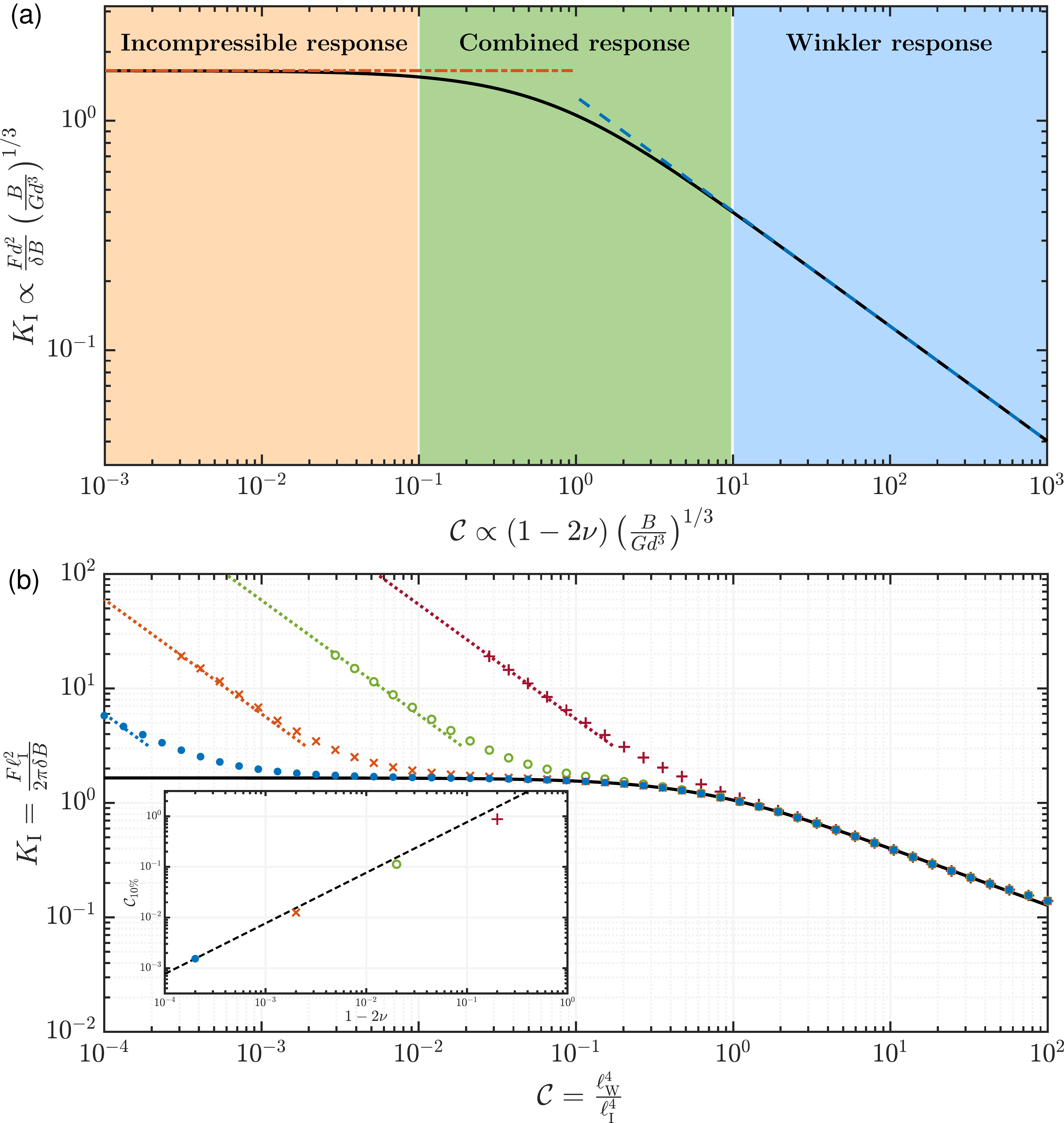}
\caption{The dimensionless indentation stiffness, $\KD\propto F/\delta$, for a stiff coating  of a thin elastomeric layer with varying compressibility $\C$. (a) The prediction of the combined foundation model, eq.~\eqref{eq:stiffcoat_fullsol_stiff}, recovers the result for a Winkler mattress, \eqref{eq:WinkIndStiffness} (dashed line) in the compressible limit, $\C\to\infty$, and that for an incompressible foundation, \eqref{eq:DillIndStiffness}, (dash-dotted line) in the incompressible limit, $\C\to0$.  Shading illustrates the relevant substrate response in each range of values of $\C$. (b) Main figure: Comparison between the prediction of the combined foundation model (solid curve) and numerical solution of the full problem  [eqs.~\eqref{eq:appStiff_equil}--\eqref{eq:appStiff_plate}] (points). Numerical results are shown for different values of the Poisson  ratio $\nu$ as follows: $\nu=0.4$ ($+$), $\nu=0.49$ ($\circ$), $\nu=0.499$ ($\times$) and $\nu=0.4999$ ($\bullet$). We observe good agreement between the numerics and the combined foundation model provided that $\C\gg(1-2\nu)$. For $\C\lesssim 1-2\nu$, our numerical results tend to the analytical result for an infinitely deep substrate, \eqref{eqn:BoxStiffness} (dotted lines). Inset: The $\C$ for which the  relative error in the use of the combined foundation model is  10\%, denoted $\C_{10\%}$, scales linearly with $1-2\nu$ (dashed line); for $\C\geq\C_{10\%}$ the results of the combined foundation model lie within $10\%$ of the numerical results.
\label{fig:IndentationStiffness}}
\end{figure}

Figure \ref{fig:IndentationShapes} shows the surface profile predicted by \myeqref{eq:stiffcoat_fullsol}{a} for three different values of $\C$. (Since different values of $\C$ correspond to the incompressible or Winkler characters of the substrate dominating, this figure shows each profile plotted with respect to the different horizontal length scales $\lD$ and $\lW$ on the left and right, respectively.) The behaviour of the indentation stiffness, $\KD$, as a function of $\C$, as given in \eqref{eq:stiffcoat_fullsol_stiff}, is plotted in Fig.~\myref{fig:IndentationStiffness}{a}. 

To further understand the behaviour of the solution in \eqref{eq:stiffcoat_fullsol}, we  reconsider the two limits of interest: the incompressible limit ($\C\to 0$) and the Winkler/compressible limit ($\C\to \infty$).

\subsubsection{Incompressible foundation, $\C\to 0$}

To recover the incompressible foundation behaviour from \eqref{eq:stiffcoat_fullsol} we let $\C\to 0$, leading to $\lambda\sim 1$ and $\Lambda\sim e^{\im\pi/3}$. The deflection profile may therefore be written (in dimensional form) as
\begin{subequations}\label{eq:IndDill_full}
\begin{equation}\label{eq:IndDillProf}
\frac{\zeta(r)}{\delta} \sim  \frac{\sqrt{3}}{\pi}\Re\left\{K_0\left[\frac{r}{\lD}\right]-2e^{\im\pi/3}K_0\left[\frac{r}{\lD}e^{\im\pi/3}\right]\right\},
\end{equation}
where $\delta$ is related to the applied load $F$ by the stiffness
\begin{equation}\label{eq:DillIndStiffness}
\frac{F}{\delta}\sim 6\sqrt{3}\frac{B}{\lD^2}=6\sqrt{3}\left(\frac{B^2 G}{\Gamma(\nu)d^3}\right)^{1/3}.
\end{equation}
\end{subequations}
While the related case of  line-indentation of a plate on top of a thin, perfectly incompressible,  foundation has been  studied previously \cite{Dillard1989, Bert1994, Dillard2018, Mukherjee2018}, to our knowledge, the axisymmetric solution \eqref{eq:IndDill_full} is new.

\subsubsection{Winkler foundation, $\C\to \infty$}

The Winkler foundation behaviour is recovered  in the limit $\C\to \infty$, for which $\lambda\sim \C^{1/2}$ and $\Lambda\sim e^{\im\pi/4}\C^{-1/4}$. The deflection profile may therefore be written (in dimensional form) as
\begin{subequations}
\begin{equation}\label{eq:IndWinkProf}
\frac{\zeta(r)}{\delta} \sim  \frac{4 }{\pi} \Im\left\{K_0\left[\frac{r}{\lW} e^{\im\pi/4}\right]\right\} \equiv  \frac{4 }{\pi} \mathrm{kei}_0\left[\frac{r}{\lW}\right],
\end{equation}
where $\mathrm{kei}_0(x)$ is the Kelvin kei function of zeroth-order \cite{Abramowitz1964}; now $\delta$ is related to $F$ by
\begin{equation}\label{eq:WinkIndStiffness}
\frac{F}{\delta}\sim 8\frac{B}{\lW^2}= 8\left(\frac{B G}{\alpha(\nu)(1-2\nu)d}\right)^{1/2},
\end{equation}
\end{subequations} which is  equivalent to the result of Hertz \cite{Hertz1884} for the response of a plate floating on a liquid bath. (In Hertz' problem  the hydrostatic pressure within the liquid is linear in deflection, and hence behaves as a perfect Winkler foundation \cite{Dempsey1991, Wang2005, Box2017}.)
 
The results presented in Fig.~\myref{fig:IndentationStiffness}{a} show that the combined foundation model interpolates between the incompressible and Winkler limits. In particular, we find that the incompressible foundation is recovered for $\C\lesssim 10^{-1}$ while the Winkler foundation model is recovered for $\C\gtrsim 10^{1}$. For $10^{-1}\lesssim \C \lesssim 10^1$ the substrate does not sit cleanly in either limit and the combined foundation model, \eqref{eqn:CombFound}, must be used.
 
\subsubsection{Model validity}

The analysis presented so far rests on the assumption that the length scale $\ell$ is large enough to justify the thin layer analysis of the substrate deformation \emph{and} the use of the plate equation. Together, these assumptions amount to assuming that
\begin{equation}\label{eq:stiffcoat_fail}
\max\{d^2, t^2\}\ll \ell^2.
\end{equation} Since the appropriate value of $\ell$   depends  on $\C$, we divide both sides by $d^2$ and note that $(\lD/d)^2\equiv \C\Gamma(\nu) /(1-2\nu)\alpha(\nu)$ and  $(\lW/d)^2\equiv  \C^{3/2}\Gamma(\nu) /(1-2\nu)\alpha(\nu)$, so that we can rewrite \eqref{eq:stiffcoat_fail} as: 
\begin{equation}\label{stiffcoat_limits}
\max\{1, \that^2\}\ll \frac{\Gamma(\nu)}{(1-2\nu)\alpha(\nu)}\max\{\C,\C^{3/2}\},
\end{equation}
where $\that\coloneqq t/d$ is  the ratio of  sheet thickness to the substrate depth.

The limit on the size of $\that$  in \eqref{stiffcoat_limits} expresses when modelling  the stiff coating as a Kirchhoff--Love plate is valid. Of more interest here, however, is the first inequality, which describes the validity of the small foundation aspect ratio  assumption ($\epsilon^2\equiv d^2/\ell^2\ll 1$) that was used to derive \eqref{eq:Chandler_found}. In particular, for the combined foundation model to be valid, we require that:
\begin{equation}\label{eq:stiffcoat_faillimit}
1\ll\frac{\Gamma(\nu)}{(1-2\nu)\alpha(\nu)}\max\{\C,\C^{3/2}\}\equiv\max\left\{\left[\Gamma(\nu)\frac{B}{G d^3}\right]^{1/3},\left[(1-2\nu)\alpha(\nu)\frac{B}{G d^3}\right]^{1/2}\right\}.
\end{equation}
To understand this condition better, we compare the predictions of the combined foundation model with numerical solutions of the full equilibrium equations  \eqref{eq:Navier_eq} for the substrate.

\subsection{Comparison with numerical results }\label{sec:Numerical_comp}

When the assumption of small aspect ratios ($\epsilon\ll 1$) is no longer valid, one must instead solve the full two-dimensional equilibrium equations \eqref{eq:Navier_eq},  on $-d \leq z\leq 0$ and $0\leq r<\infty$, subject to the boundary conditions \eqref{eq:Navier_bcs} with $\bm{T}(\bm{x})$ chosen such that $u(r,0)=0$ and $p(\bm{x})\equiv q(r)$ given by the plate equation \myeqref{eq:stiffcoat_eqs}{a} and \eqref{eq:stiffcoat_bcs}. Non-dimensionalizing using the scalings presented in \eqref{eq:stiffcoat_dimvars} and noting that the aspect ratio can be written as
\begin{equation}\label{eq:stiffcoat_eps_Nnu}
\epsilon^2=\frac{d^2}{\lD^2} \equiv \frac{\alpha(\nu)(1-2\nu)}{\Gamma(\nu)\C},
\end{equation}
we obtain a system controlled by the Poisson ratio $\nu$ and the compressibility $\C$, defined in \eqref{eq:stiffcoat_N}. In the full problem, the Poisson ratio can no longer be scaled out of the problem (as was possible for $\epsilon\to0$); $\nu$ is an additional dimensionless parameter.  The resulting system of partial differential equations may be solved using standard numerical techniques; here we use the method of finite differences. The full system and further numerical discussion is presented in \ref{app:Numerics_example1}. 

For given values of the parameters $\nu$ and $\C$, we can compute the indentation stiffness using the first integral of  \myeqref{eq:stiffcoat_eqs_nondim}{a} for any foundation aspect ratio, \eqref{eq:stiffcoat_eps_Nnu}. Figure \myref{fig:IndentationStiffness}{b} shows the dimensionless stiffness, $\KD$, for varying $\C$ and $\nu$ as computed numerically, together with the prediction of the combined foundation model \eqref{eq:stiffcoat_fullsol_stiff}, and the corresponding result for an infinitely deep foundation \eqref{eqn:BoxStiffness} \cite{Box2020}. This plot shows excellent  agreement between the combined foundation model and the numerics for sufficiently large $\C$ (or, equivalently, sufficiently small aspect ratios, $\epsilon\ll1$). For small $\C$ (or, equivalently, large aspect ratios, $\epsilon\gg1$), however, we see that the numerical results are better predicted by \eqref{eqn:BoxStiffness} for an infinitely deep foundation. Crucially, the domain of validity of the combined foundation model  reaches  smaller $\C$ as $\nu\to1/2$, as expected from \eqref{eq:stiffcoat_faillimit}: in particular, the regime for which the results of the combined foundation model lie within $10\%$ of the numerical model, $\C\geq\C_{10\%}$ grows as $\nu\to1/2$ (see inset of fig.~\ref{fig:IndentationStiffness}b).

In the example considered in this section, the horizontal length scale was set simply by the compressibility of the substrate and was, hence, a material parameter, independent of, for example, the applied load (\ie~$F$). We now consider a more involved example in which the length scale itself varies with the applied load.
%-----------------------------------------------------------------------

\section{Second example: A lubrication problem}\label{sec:SecondExample}

A classic problem in fluid mechanics is the ability to lubricate two solid bodies using a thin layer of a viscous liquid  \cite{Jeffrey1981}.  Lubrication relies on tangential motion generating a vertical force that can support load; given this ubiquity, it is at first surprising that with rigid, fore-aft symmetric boundaries,  viscous fluid flow cannot generate a significant vertical lift force --- a consequence of Purcell's so-called `Scallop Theorem' \cite{Purcell1977,Lauga2011}. There are many means by which this symmetry can be broken (including cavitation \cite{Ashmore2005}, non-Newtonian effects and inertia \cite{Lauga2011}) but one of the most common is the inclusion of a deformable  boundary \cite{Lauga2011}. Since the coupling between elastic deformation and fluid flow can make analytical work difficult,  a popular model of a deformable boundary is the Winkler model;  several  works have studied the coupling between a Winkler mattress and the  flow induced by a horizontally translating cylinder \cite{Skotheim2004,Skotheim2005,Salez2015} or sphere \cite{Weekley2006,Davies2018}.

Several  recent studies have focussed on experimental realizations of the sedimentation of a rigid cylinder in a viscous liquid above an inclined plane with a soft, elastomeric coating \cite{Saintyves2016, Saintyves2020,Rallabandi2017}. Such an experiment is conducted under  force-controlled conditions: the cylinder sediments until the separation between the substrate and its base generates sufficient hydrodynamic force to balance the cylinder's weight; at this point, the cylinder translates at a constant height above the surface of the substrate, and may also rotate \cite{Saintyves2016}. The theoretical understanding of these experiments \cite{Rallabandi2017,Saintyves2020} generally rests on the predictions of Winkler based models, while experiments use elastomeric coatings that are (close to) incompressible \cite{Saintyves2016}. Given the preceding discussion, it is natural to ask whether/when such experiments operate in the compressible or incompressible regimes? In this section, we use the combined model \eqref{eqn:CombFound} to address this question.

\subsection{Problem setup}

\begin{figure}[ht!]
\centering
\includegraphics[width=.8\textwidth]{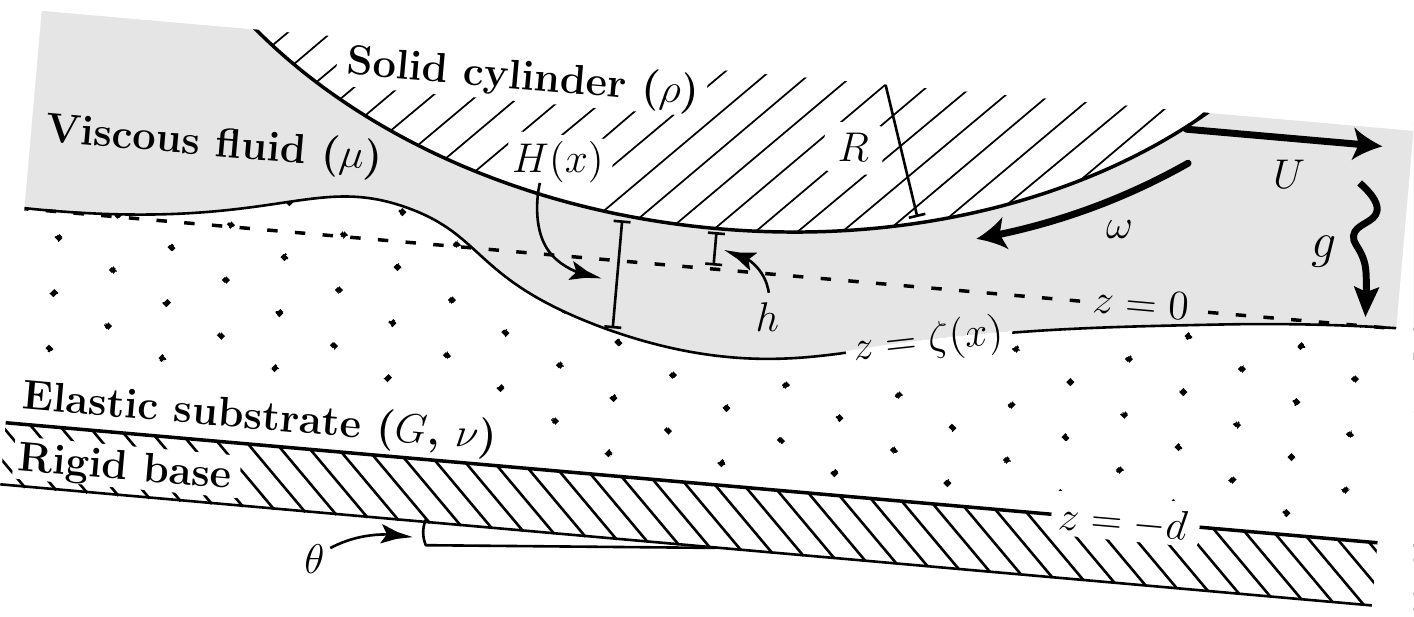}
\caption{A sketch of the second example: a two-dimensional cylinder (of radius $R$) translates and rotates above an inclined elastic substrate. The cylinder's weight is supported  by the lift force that is generated by the interaction between a viscous fluid flow and substrate deformation. The resulting gap profile is denoted $H(x)=h+x^2/2R -\zeta(x)$ where $\zeta(x)$ is the  profile of the deformed surface  and  $h$ is the minimum gap thickness without deformation. 
\label{fig:CylSetup}}
\end{figure}

We consider a two-dimensional cylinder (of radius $R$ and density $\rho$) translating perpendicular to its axis at a uniform speed $U$, and rotating clockwise about its axis at an angular speed $\omega$, in a viscous fluid (of viscosity $\mu$) suspended above a thin elastic substrate  (of depth $d$, Poisson ratio $\nu$, and shear modulus $G$), which is inclined at an angle $\theta$ to the horizontal, as shown schematically in Fig.~\ref{fig:CylSetup}. The motion of the cylinder parallel to the foundation forces liquid to flow beneath the cylinder and out again, introducing a one-dimensional pressure field $p(x)$ (measured with respect to atmospheric pressure) that is positive ahead of the cylinder and negative behind it. (Note that here, negative pressures are relative to atmospheric and so do not, in general, correspond to cavitation of the liquid.) The deformation of the substrate introduces an asymmetry in $p(x)$ that has a non-zero integral and hence is able to provide a sustained lift to support the cylinder. The magnitude of this lift will depend on the minimum separation between the cylinder and the undeflected surface of the soft substrate, which we denote by $h$.

The parabolic geometry near the minimum cylinder--substrate separation means that lubrication problems of this type have a characteristic horizontal length scale $\ell\propto (hR)^{1/2}$ \cite{Jeffrey1981}. However, $h$ is itself not known \emph{a priori} --- it is determined by the requirement to generate enough lift force to support the cylinder's weight, which in turn depends on the type of substrate response (Winkler or incompressible), and hence $h$ depends on $\ell$. To understand this coupling better, we study the behaviour in terms of the experimental control parameters (the density of the cylinder \etc), rather than emergent physical properties (such as $h$). To do this, we use the combined foundation model [\ie~taking $p(\bm{x})\mapsto p(x)$ in \eqref{eq:Chandler_found}] coupled with Reynolds' equation \cite{Leal2007} to describe the fluid flow in the narrow gap beneath the cylinder. In particular, in the reference frame of the translating cylinder, we have
\begin{subequations}\label{eq:viscyl_eqs}
\beq
\frac{\de\,}{\de x}\left[H(x)^3\frac{\upd p}{\upd x} + 6 \mu \left(U+\omega R\right) H(x)\right]=0,
\label{eq:viscyl_eqs_Rey}
\eeq
where the separation between the cylinder surface and  that of the elastomeric coating is
\beq
H(x)= h+\frac{x^2}{2 R} -\zeta(x).
\eeq
The  displacement of the elastomer surface, $\zeta(x)$, is given in terms of the pressure field by \eqref{eq:Chandler_found}, \ie
\beq
\zeta(x)= -\alpha(\nu)(1-2\nu)\frac{d}{G} p(x)+ \Gamma(\nu)\frac{d^3}{G} \frac{\upd^2p}{\upd x^2},
\label{eq:viscyl_eqs_Disp}
\eeq
\end{subequations} where we neglect the  effect of the viscous shear stress on the elastomer surface; this shear stress is $T(x)\sim H(x) p'(x)$, so provided $[H]\ll d$ this neglect of $T$ is appropriate and we may use the expression for $\Gamma(\nu)$ for zero surface-shear in \eqref{eq:fuctions_fg}. Solutions of \eqref{eq:viscyl_eqs} are subject to far-field decay conditions,
\begin{equation}\label{eq:viscyl_bcs}
p(x) \to 0 \qquad\text{as}\quad x\to \pm \infty.
\end{equation}

In experiments \cite{Saintyves2016,Saintyves2020},  the gap thickness, $h$,  translational speed, $U$, and angular velocity, $\omega$, are  determined by requiring no net force perpendicular or parallel to the plane and no net torque; these conditions may be written \cite{Skotheim2005,Rallabandi2017}
\begin{subequations}\label{eq:viscyl_int}
\begin{align}
\pi R^2 \rho g \cos\theta &= \int_{-\infty}^\infty p(x)\de x,\label{eq:viscyl_int_lift}\\
\pi R^2 \rho g \sin\theta &= \int_{-\infty}^\infty\left[ \frac{x}{R}p(x)+ \left.\tau_{xz}\right|_{z=H}\right] \de x,\\
0 &=\int_{-\infty}^\infty R \left.\tau_{xz}\right|_{z=H}\de x=R\int_{-\infty}^\infty \left[\frac{1}{2}H(x)p'(x)+\frac{\mu(U-\omega R)}{H}\right]\de x,
 \end{align}
 \end{subequations}
respectively. Note that \myeqref{eq:viscyl_int}{c} immediately eliminates the second term in the integrand of   \myeqref{eq:viscyl_int}{b}.

\subsubsection{Scaling analysis}

A key difference between the present problem and that presented in \S\ref{sec:FirstExample} is that the pressure scale that determines the deformation of the elastomeric foundation is not clearly related to the pressure that supplies the normal force to balance the weight of the cylinder. This is because the pressure field associated with the leading-order problem (the rigid case) is anti-symmetric, and hence does not supply a normal load \cite{Skotheim2004}. This leading-order problem does, however, dominate the tangential force balance \myeqref{eq:viscyl_int}{b}, which, when combined with a scaling analysis of Reynolds' equation \eqref{eq:viscyl_eqs_Rey},  allows us to determine the leading-order pressure and velocity scales, $p_*$ and $U_*\sim R \omega_*$, respectively. In particular, assuming that the steady translation of the cylinder occurs with a separation $h$ from the undisturbed level of the substrate, and hence that the relevant horizontal scale is $\propto(hR)^{1/2}$, Reynolds' equation \eqref{eq:viscyl_eqs_Rey} predicts  that
\beq
p_\ast\propto\frac{\mu U_\ast R^{1/2}}{h^{3/2}}\sim\frac{\mu \omega_\ast R^{3/2}}{h^{3/2}},
\eeq
 while \myeqref{eq:viscyl_int}{b} predicts that $\rho g R^2\sin\theta\propto hp_\ast$. Solving these two equations gives
\begin{subequations}\label{eqn:DVestimates}
\begin{equation}
p_\ast\propto\frac{\rho g R^2\sin\theta}{h}\quad \mathrm{and}\quad U_\ast \sim R\omega_*\propto\frac{\rho g R^{3/2}h^{1/2}\sin\theta}{\mu},\subtag{a,b}
\end{equation}
\end{subequations}
 with $h$ still undetermined. As already discussed, $h$ must depend on the deformability of the substrate: without substrate deformation, no value of $h$ can generate an equilibrium lift force.  Nevertheless, we can use the scaling estimates \eqref{eqn:DVestimates} to estimate the deflection of the foundation  for each of the Winkler and incompressible limits, denoted $\zeta_\ast$; we find that
\beq
\zeta_\ast \propto \begin{dcases} (1-2\nu)\frac{d}{G}p_\ast \qquad&\text{(Winkler)},\\
\frac{d^3}{G}\frac{p_\ast}{hR} \qquad&\text{(incompressible)}.
\end{dcases}
\label{eqn:ZetaScales}
\eeq

The final piece of the scaling puzzle  is to understand the effect of the deflection $\zeta_\ast$ on the pressure field; this perturbation to the pressure field, $\delta p$, is what supports the cylinder's load. Since the pressure depends on gap thickness algebraically, the relative change in pressure should be proportional to the relative change in gap width, \ie~$\delta p/p_\ast \propto \zeta_*/h$. Vertical force balance then gives $\rho g R^2\cos\theta \propto \delta p (hR)^{1/2} \propto p_\ast \zeta_\ast(R/h)^{1/2}$ and  two possible scalings for the separation $h$, depending on which of the deflection scales in \eqref{eqn:ZetaScales} is more appropriate:
\beq
h \propto\begin{dcases} \left(\frac{(1-2\nu)\rho g d\sin\theta\tan\theta}{G}\right)^{2/5}R\qquad&\text{(Winkler),}\\
\left(\frac{\rho g d^3\sin\theta\tan\theta}{G}\right)^{2/7}R^{3/7}\qquad&\text{(incompressible)}.
\end{dcases}
\label{eqn:h0Scales}
\eeq 
These scalings have been given before by Skotheim \& Mahadevan \cite{Skotheim2004,Skotheim2005}; our aim here is to go beyond the scaling analysis and investigate the transition between the Winkler and incompressible regimes. We therefore turn first to the non-dimensionalization of the problem, which will be informed by the scaling behaviour determined above.

\subsubsection{Non-dimensionalization}

From the preceding scaling analysis, there are two limiting scalings for the  horizontal length $\ell$:
\begin{subequations}\label{eq:viscyl_LDLW}
\begin{equation}
\lW \coloneqq d\left( \alpha(\nu)(1-2\nu)\frac{ 2R^5\rho g \sin\theta\tan\theta}{d^4G}\right)^{1/5} \text{ or}\enskip 
\lD \coloneqq d\left(\Gamma(\nu)\frac{ 2R^5 \rho g \sin\theta\tan\theta}{d^4G}\right)^{1/7},
\subtag{a,b}
\end{equation}
\end{subequations}
which correspond to foundations whose response is of the Winkler or incompressible type, respectively. Without loss of generality, we choose the incompressible length scale, $\lD$, as the  horizontal scale and,  motivated by the preceding scaling analysis, define vertical scales $[h]=\lD^2/(2R)$ and $[\zeta]\coloneqq \lD^3/(2 R^2 \tan\theta)$. (We emphasize that these scales are defined in terms of experimental control parameters, rather than the emergent properties $h$, $U$, and $\omega$.)

We  introduce the dimensionless variables
\begin{subequations}\label{eq:viscyl_dimvars}
\begin{equation}
 X \coloneqq  \frac{x}{\lD}, \quad Z(X)\coloneqq \frac{2R^2\tan{\theta}}{\lD^3}\zeta(x), \quad \Hcal(X)\coloneqq \frac{2R  }{\lD^2}H(x), \quad P(X) \coloneqq \frac{\lD^2 }{R^3 \rho g \sin\theta} p(x).
\subtag{a--d}
\end{equation}
\end{subequations}
Inserting \eqref{eq:viscyl_dimvars} into the model, \eqref{eq:viscyl_eqs}, we find the dimensionless system
\begin{subequations}\label{eq:viscyl_sys_dimless}
\begin{gather}
0=\frac{\de\,}{\de X}\left[ \Hcal(X)^3  P'(X)+ 6(\Uhat +\ohat)\Hcal(X)\right],\\
\Hcal(X) = \hhat+X^2 -\phhat Z(X),\qquad
Z(X) = -\Chat P(X)+  P''(X), \subtag{b,c}
\end{gather}
which is to be solved subject to $P(X) \to 0$ as $X\to \pm \infty$ and the integral constraints \eqref{eq:viscyl_int}, which after simplification read:

\begin{gather}\label{eq:viscyl_integral_dimless}
\phhat  = \frac{1}{\pi}\int_{-\infty}^\infty P(X)\de X, 
\qquad
1 = \frac{1}{\pi}\int_{-\infty}^\infty X P(X)\de X, \subtag{d,e}\\
0 =\int_{-\infty}^\infty \frac{1}{2}\Hcal(X)P'(X)+\frac{\Uhat-\ohat}{\Hcal(X)}\de X. \subtag{f}
\end{gather}
\end{subequations} 
Note that Reynolds' equation with $\Hcal(x)=\Oh(X^2)$ as $|X|\to \infty$ gives $P(X)=\Oh(1/X^3)$ as $|X|\to  \infty$; using this and the combined foundation \myeqref{eq:viscyl_sys_dimless}{c} gives $\int_{-\infty}^\infty\Hcal(X)P'(X)/2\de X=-\int_{-\infty}^\infty X P(X)\de X$, allowing further simplification of \myeqref{eq:viscyl_sys_dimless}{f}.

The  system \eqref{eq:viscyl_sys_dimless} is  governed by the, experimentally-controlled, dimensionless parameters
\begin{subequations}\label{eq:viscyl_NphiUh}
\begin{equation}
\Chat \coloneqq  \frac{ \alpha(\nu)(1-2\nu)}{\Gamma(\nu)} \left(\frac{\lD}{d}\right)^2\quad \text{and} \quad \phhat \coloneqq \frac{\lD}{R\tan\theta} ,
\subtag{a,b}
\end{equation}
but also involve three emergent quantities, the dimensionless gap thickness, cylinder speed, and rotation speed, $\hhat$, $\Uhat$, and $\ohat$, respectively --- these are to be determined as part of the solution. With these rescalings, the physical minimum gap thickness, $h$, translational speed, $U$, and rotational speed, $\omega$, are given by
\begin{equation}
h=\hhat \times \frac{\lD^2}{2R}, \qquad U=\Uhat \times \frac{\lD R \rho g \sin\theta}{4\mu}, \qquad  \omega=\ohat \times \frac{\lD \rho g \sin\theta}{4\mu}.
\subtag{c--e}
\end{equation}
\end{subequations}

We also note that the dimensionless parameter $\Chat\equiv(\lW/\lD)^5\propto (1-2\nu)$: $\Chat$ is a measure of the compressibility of the substrate, analogous to the parameter $\C$ in the previous example \eqref{eq:stiffcoat_N}. In addition, $\phhat\equiv[\zeta]/[H]$  so that $\phhat$ is a measure of how pliable the substrate is.  In experiments $\phhat$ is often  small [for example in \cite{Saintyves2016, Saintyves2020}  $\phhat=\Oh(10^{-2}\mbox{--}10^{-1})$] --- the elastic substrate is relatively stiff and makes only small departures from its undeformed planar state. In the limit $\phhat\to 0$ it is possible to make analytical progress and so we turn first to this limit before considering numerical solutions.

\subsection{Analytical Results for $\phhat\ll1$}\label{sec:SecondExample_Analytical}

Following the  perturbative approach of Skotheim \& Mahadevan \cite{Skotheim2004, Skotheim2005} for a Winkler foundation, we pose the regular expansions
\begin{subequations}
\begin{gather}
Z(X)=Z_0(X)+\phhat Z_1(X)+\Oh(\phhat^2),\qquad
P(X)=P_0(X)+\phhat P_1(X)+\Oh(\phhat^2),\subtag{a,b}\\
\hhat=\hhat_0+\phhat \hhat_1+\Oh(\phhat^2),\qquad
\Uhat=\Uhat_0+ \phhat \Uhat_1+\Oh(\phhat^2),\qquad
\ohat=\ohat_0+\phhat \ohat_1+\Oh(\phhat^2),\subtag{c--e}
\end{gather}
\end{subequations}
in \eqref{eq:viscyl_sys_dimless}, and consider the problem order-by-order. Note that our definition of the gap separation $\Hcal(X)$ in \myeqref{eq:viscyl_sys_dimless}{b} included a factor of $\phhat$ multiplying $Z(X)$ --- hence the deflection \emph{does} return to the rigid parabolic geometry as $\phhat\to0$.

At leading-order, $\Oh(\phhat^0)$, we recover the rigid-substrate  problem \cite{Jeffrey1981}, as  expected; this determines the leading-order surface deflection, $\phhat Z_0(X)$, which is dependent on $P_0(X)$ as:
\begin{subequations}\label{eq:viscyl_asymp_profiles_PZ}
\begin{align}
P_0(X) &= \frac{2(\Uhat_0+\ohat_0) X}{(\hhat_0+X^2)^2},\label{eq:viscyl_asymp_profiles_P}\\
Z_0(X) &= -\frac{2(\Uhat_0+\ohat_0)  X}{(\hhat_0+X^2)^2}\Chat-\frac{24(\Uhat_0+\ohat_0)  X(\hhat_0-X^2)}{(\hhat_0+X^2)^4}.\label{eq:viscyl_asymp_profiles_Z}
\end{align}
\end{subequations}
Substituting these expressions into the integral constraints, \myeqref{eq:viscyl_sys_dimless}{d--f} at $\Oh(\phhat^0)$, gives
\begin{subequations}\label{eq:viscyl_order0}
\begin{equation}
\Uhat_0 = \hhat_0^{1/2} \quad \text{and} \quad \ohat_0=0.\subtag{a,b}
\end{equation}
\end{subequations}

At the next-order, $\Oh(\phhat^1)$, we obtain expressions for $P_1(X)$ and $Z_1(X)$ --- these were calculated using \textsc{Mathematica}, and are omitted here for reasons of space  --- substituting these expressions into the integral constraints, \myeqref{eq:viscyl_sys_dimless}{d--f} at $\Oh(\phhat^1)$, gives
\begin{subequations}\label{eq:viscyl_order1}
\begin{equation}
\hhat_0^{7/2} = \frac{3}{8}\Chat\hhat_0+\frac{45}{16},\qquad \Uhat_1 = \frac{\hhat_1}{2\hhat_0^{1/2}}, \qquad \ohat_1=0.\subtag{a--c}
\end{equation}
\end{subequations}
Solving \myeqref{eq:viscyl_order0}{a} and \myeqref{eq:viscyl_order1}{a} gives the leading-order gap thickness, $\hhat\sim \hhat_0$, and translational speed, $\Uhat\sim \Uhat_0=\hhat_0^{1/2}$. (Note that for general values of $\Chat$,  \myeqref{eq:viscyl_order1}{a} is an algebraic equation for $\hhat_0$ that must be solved numerically, although asymptotic results are readily available in the limits $\Chat\ll1$ and $\Chat\gg1$.)  Determining the leading-order rotational speed $\ohat$, however, requires that we continue to the next-order.

At the next-order, $\Oh(\phhat^2)$, we obtain expressions for $P_2(X)$ and $Z_2(X)$ --- which are again omitted here. Substituting these into the integral constraints, \myeqref{eq:viscyl_sys_dimless}{d--f} at $\Oh(\phhat^2)$, gives
\begin{subequations}\label{eq:viscyl_order2}
\begin{equation}
\hhat_1= 0 \quad \text{and} \quad \ohat_2=\frac{21\Chat^2 \hhat_0^2+462 \Chat \hhat_0+3042}{256 \hhat_0^{11/2}}.\subtag{a,b}
\end{equation}
\end{subequations}
Using the solution of \myeqref{eq:viscyl_order1}{a}, \myeqref{eq:viscyl_order2}{b} gives the leading-order rotational speed $\ohat\sim\phhat^2 \ohat_2$.

\begin{figure}[ht!]
\centering
\includegraphics[width=0.8\textwidth]{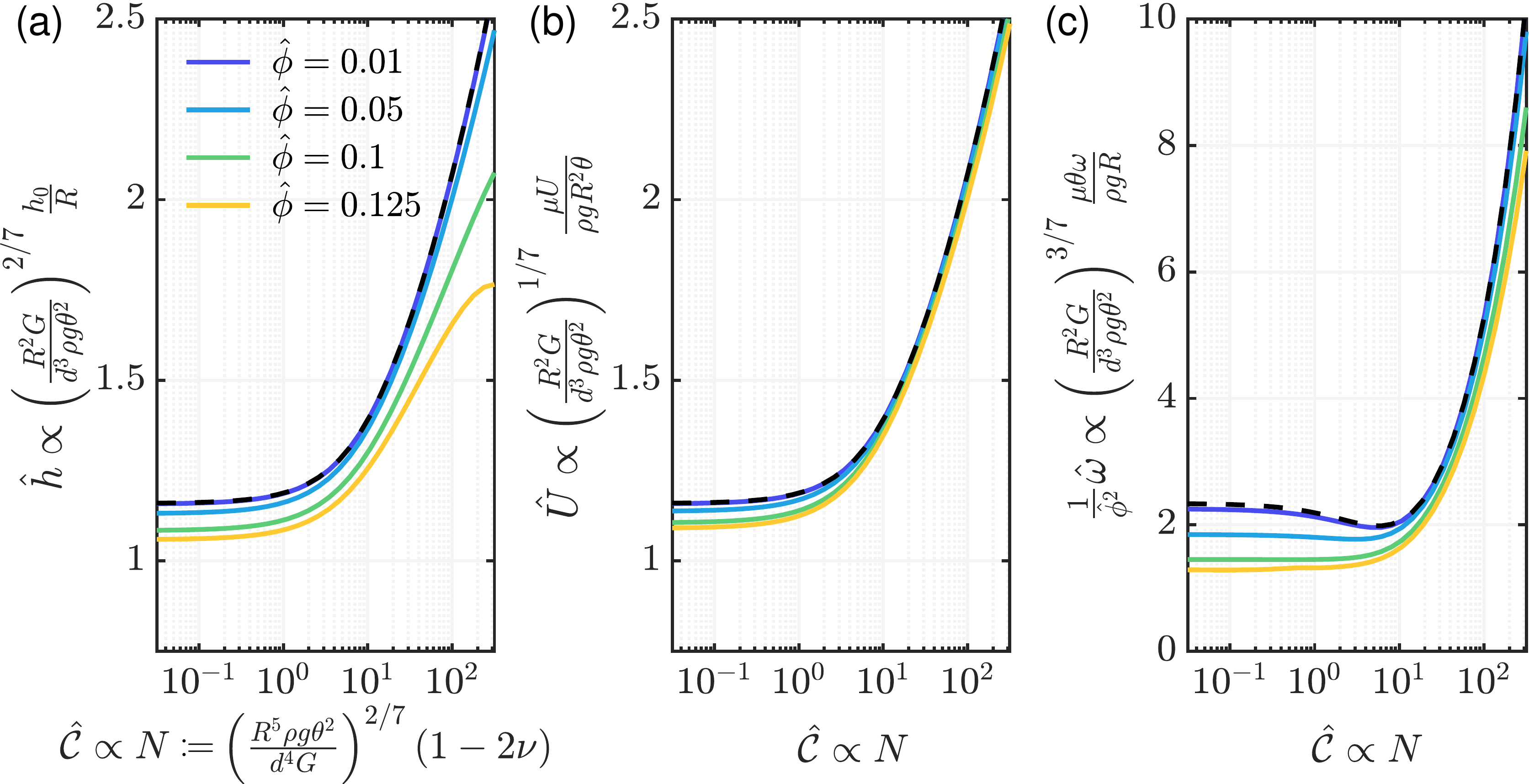}
\caption{Comparison of asymptotic (valid for small pliability, $\phhat\ll1$) and numerical results of the combined foundation model applied to a cylinder sedimenting close to a thin, elastomeric layer of dimensionless compressibility $\Chat$. (a) The minimum gap thickness, (b) steady translation speed, and (c) steady rotation speed. In each figure, the asymptotic prediction \eqref{eq:viscyl_order0}--\eqref{eq:viscyl_order2}, valid for $\phhat\ll1$, is shown by dashed curves while solid curves represent the numerical solution of the combined foundation problem with different values of the pliability parameter  $\phhat$, as given in the legend. Here we have  assumed  $\theta\ll1$ to simplify the axes labels.
\label{fig:CylCompFound}}
\end{figure}

The full asymptotic behaviour (as $\phhat\to0$)  for varying substrate compressibility, $\Chat$, is plotted in Fig.~\ref{fig:CylCompFound} together with numerical solutions for non-zero $\phhat$ (see \S\ref{sec:SecondExample}\ref{sec:LubricationNums} below for details). These numerical results demonstrate that \eqref{eq:viscyl_order0}--\eqref{eq:viscyl_order2} recover the correct behaviour for small $\phhat$.

To  understand the implications of the above asymptotic results better, we move now to  discuss the behaviour in the incompressible limit ($\Chat\to 0$, incompressible foundation) and the limit of high compressibility ($\Chat\to \infty$, Winkler foundation) in more detail.

\subsubsection{Incompressible foundation, $\Chat\to0$}

For an incompressible  foundation,  $\Chat\to 0$,  we immediately find from \eqref{eq:viscyl_order0}--\eqref{eq:viscyl_order2} that $\hhat\sim (45/16)^{2/7}$, $\Uhat\sim(45/16)^{1/7}$, and $\ohat\sim 338/225 \times (45/16)^{3/7}\phhat^2$ so that the dimensional gap thickness, translational speed, and rotational speed are
\begin{subequations}\label{eqn:IncompScalings}
\begin{gather}
h\sim\frac{\lD^2}{2R}\times \left(\frac{45}{16}\right)^{2/7}, \qquad
U\sim\frac{ R\lD \rho g \sin\theta}{4\mu}\times \left(\frac{45}{16}\right)^{1/7} , \subtag{a,b} \\
\omega\sim \frac{\rho g \lD^3\cos\theta}{4\mu R^2\tan\theta}\times\frac{338}{225}\left(\frac{45}{16}\right)^{3/7}. \subtag{c}
\end{gather}
\end{subequations}
In this incompressible limit, we also have the pressure and surface-displacement profiles
\begin{subequations}
\begin{align}
P(X)&\sim\frac{2\hhat_0^{1/2} X}{(\hhat_0+X^2)^2} +\frac{48(11\hhat_0^2-28\hhat_0 X^2+21X^4)}{35(\hhat_0+X^2)^7}  \hhat_0\phhat,\\ 
 Z(X)&\sim-\frac{24\hhat_0^{1/2}
X(\hhat_0-X^2)}{(\hhat_0+X^2)^4}.\label{eq:viscyl_dillprof}
\end{align}
\end{subequations} 
The displacement profile \eqref{eq:viscyl_dillprof}  is plotted as the solid curve in Fig.~\myref{fig:CylIntShapes}{a} and illustrates that this profile has  four turning points, located at $X=\pm\{\hhat_0(1\pm 2/\sqrt{5})\}^{1/2}=\pm(45/16)^{1/7}(1\pm 2/\sqrt{5})^{1/2}$.  (Although previous studies on the scalings of the incompressible limit have been made \cite{Skotheim2004,Skotheim2005,Rallabandi2017}, both the prefactors in \eqref{eqn:IncompScalings} and the displacement profile \eqref{eq:viscyl_dillprof} are believed new.)

\subsubsection{Winkler foundation, $\Chat\to\infty$}

The case of a Winkler foundation, the limit $\Chat\to \infty$, has been studied extensively previously \cite{Skotheim2004, Skotheim2005, Salez2015, Rallabandi2017}. In this limit, we find from  \eqref{eq:viscyl_order0}--\eqref{eq:viscyl_order2} that $\hhat\sim (3\Chat/8)^{2/5}$, $\Uhat\sim(3\Chat/8)^{1/5}$, and $\ohat\sim 7/12 \times (3\Chat/8)^{3/5}\phhat^2$ so that the dimensional gap thickness, translational speed, and rotational speed are
\begin{subequations}\label{eq:viscyl_winkhuo}
\begin{gather}
h\sim\frac{\lW^2}{2R} \times \left(\frac{3}{8}\right)^{2/5}, \qquad 
U\sim\frac{ R\lW\rho g \sin\theta}{4\mu} \times  \left(\frac{3}{8}\right)^{1/5}, \subtag{a,b}\\
\omega\sim \frac{\rho g \lW^3\cos\theta}{4\mu R^2\tan\theta} \times\frac{7}{12} \left(\frac{3}{8}\right)^{3/5}.\subtag{c}
\end{gather}
\end{subequations}
Moreover, we obtain the pressure and surface-displacement profiles:
\begin{subequations}
\begin{align}
P(X)&\sim\frac{2\hhat_0^{1/2} X}{(\hhat_0+X^2)^2} +\frac{3(3\hhat_0-5X^2)}{5(\hhat_0+X^2)^5}\Chat \hhat_0\phhat,\\ 
 Z(X)&\sim-\frac{2\Chat\hhat_0^{1/2} X}{(\hhat_0+X^2)^2}\label{eq:viscyl_winkprof}.
\end{align}
\end{subequations}
(Recall that $P(X)$, $Z(X)$, and $X$ are all non-dimensionalized with respect to the incompressible response, hence, there is an explicit dependence on $\Chat$ above.) The displacement profile \eqref{eq:viscyl_winkprof}, recovers that given previously \cite{Skotheim2004,Skotheim2005}; it is plotted in Fig.~\myref{fig:CylIntShapes}{a}, and, in contrast to the incompressible profile \eqref{eq:viscyl_dillprof}, has only two turning points, located at  $X=\pm(\hhat_0/3)^{1/2}=\pm(\Chat/24\sqrt{3})^{1/5}$. The gap thickness, translation speed, and rotation speed \eqref{eq:viscyl_winkhuo} agree with previous results \cite{Rallabandi2017}.

\subsection{Numerical comparison\label{sec:LubricationNums}}

The analytical progress that has been made above has exploited the doubly asymptotic limit of small substrate pliability, $\phhat\ll1$, \emph{and} small aspect ratio, $d/\ell\ll1$. To investigate when our asymptotic results are valid, we use numerical solutions of the problem to study what happens when each of these limits fail. 

\subsubsection{Moderate substrate pliability}

For moderate values of $\phhat$, but retaining  small aspect ratio, there is a nonlinear feedback between Reynolds' equation and the combined foundation response, so that the system \eqref{eq:viscyl_sys_dimless} must be solved numerically. This can be done by eliminating $Z(X)$ and $\Hcal(X)$ from \eqref{eq:viscyl_sys_dimless} to form a third-order integro-differential equation for the pressure $P(X)$ with given values of  $\Chat$ and $\phhat$ and  unknown constants $\hhat$, $\Uhat$, and $\ohat$. In practice, the system can be solved by coupling finite difference derivatives and quadrature with a standard nonlinear system solver (in our work we used \texttt{fsolve} in \texttt{\textsc{Matlab}}). Once the pressure $P(X)$  has been computed,  the displacement profile $Z(X)$ can be found using \myeqref{eq:viscyl_sys_dimless}{c}.  (An alternative approach, based on similarity solutions in the long-wavelength limit, was presented recently for large deformations of a Winkler foundation by \cite{Essink2020}.)

\begin{figure}[ht!]
\centering
\includegraphics[width=0.8\textwidth]{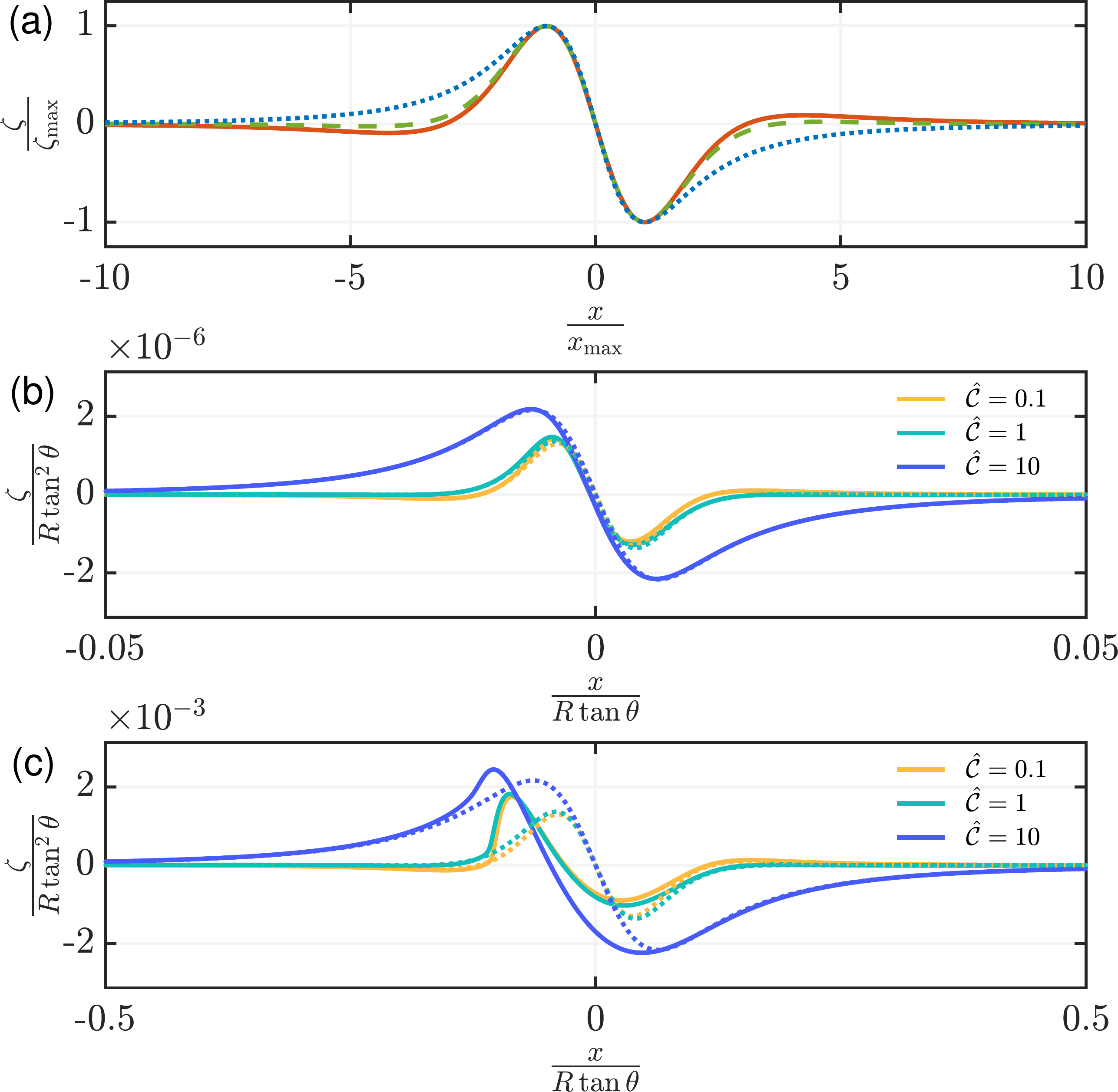}
\caption{Results of the combined foundation model for moderate substrate pliability, $\phhat$. (a) Analytical predictions for the surface deformation profile with $\phhat\ll1$, for an incompressible foundation ($\Chat\to0$) [eq.~\eqref{eq:viscyl_dillprof}], solid curve, a combined foundation ($\Chat=1$)  [eq.~\eqref{eq:viscyl_asymp_profiles_Z}], dashed curve, and a Winkler foundation ($\Chat\to\infty$) [eq.~\eqref{eq:viscyl_winkprof}],  dotted curve. To facilitate a direct comparison between the three cases, the vertical and horizontal scales have been modified to ensure that the maximum deflection occurs at $(-1,1)$. (b,c) Comparison between the predictions of the analytical results for $\phhat\ll1$ with numerical results for moderate $\phhat$ from the combined foundation model for different values of $\Chat$ (as described in the legend). In (b)~$\phhat=0.01$ and in   (c)~$\phhat=0.1$ are  fixed with numerical results  plotted as solid curves and the asymptotic predictions from eq.~\eqref{eq:viscyl_asymp_profiles_Z}  plotted as dashed curves. Note that for small $\Chat$  the additional turning points of the interface deflection predicted for an incompressible foundation (see panel a) are just visible.
\label{fig:CylIntShapes}}
\end{figure}

Numerical results for the translation velocity $\Uhat(\phhat,\Chat)$ and gap thickness $\hhat(\phhat,\Chat)$  are presented in Fig.~\ref{fig:CylCompFound} and show that the asymptotic predictions \eqref{eq:viscyl_order0}--\eqref{eq:viscyl_order2} are recovered as $\phhat\to0$ and also give a good quantitative description of the full numerical results provided that $\phhat\lesssim0.05$. Example surface profiles are presented in Fig.~\myref{fig:CylIntShapes}{b,c} for a range of $\Chat$;  these show that the asymptotic results quantitatively describe numerical results for $\phhat=0.01$, but only qualitatively for $\phhat=0.1$.

\subsubsection{Model validity}

The results presented thus far provide a mathematical description of an infinite cylinder sliding and rotating in a viscous fluid above  a thin elastic foundation. Although we have demonstrated that our asymptotic results for foundations with small pliability, $\phhat\ll1$, give good intuition for the behaviour of the combined foundation model, observed numerically for moderate $\phhat$,  these numerical results rely on the description of the substrate deformation by the combined foundation model. Now, therefore, we use numerical simulations of the full elastic-substrate problem to test when this combined foundation model is appropriate. We begin by considering the scale separation required for the model to be valid in light of the detailed analysis performed.

\paragraph{Scaling argument}
In deriving the coupled system \eqref{eq:viscyl_eqs}, we implicitly assumed that the horizontal length scale, $[x]\equiv \ell$, is large in comparison to both the coating thickness ($d^2\ll \ell^2$, for our combined foundation to be valid) and to the minimum gap thickness ($h^2\ll\ell^2$, for hydrodynamic lubrication theory to be valid). As already discussed, we also require that $h\ll d$ to be able to neglect the viscous shear force [$T(x)\sim H(x)p'(x) \ll p'(x)d$].  Altogether, we require:
\begin{equation}\label{eq:viscyl_fail}
h^2 \ll d^2\ll \ell^2.
\end{equation} 
Using the two possible horizontal length scales available in \eqref{eq:viscyl_LDLW}, we can rewrite the separation of scales required for the combined foundation model and lubrication theory to hold, \eqref{eq:viscyl_fail}, as: 
\begin{equation}
\label{eqn:LubLimits}
\left(\frac{\phhat\hhat\tan\theta}{2}\right)^2\ll \frac{\alpha(\nu)(1-2\nu)}{\Gamma(\nu)} \frac{1}{\Chat}\ll \max\{1,\Chat^{2/5}\}.
\end{equation}

The first inequality in \eqref{eqn:LubLimits} provides a parameter bound for  the fluid flow to be modelled using lubrication theory. Of more interest here, however, is the second inequality, which expresses the small foundation aspect ratio limit ($\epsilon^2\equiv \ell^2/d^2\ll 1$) that was used to derive the combined foundation model \eqref{eq:Chandler_found}. Our asymptotic results in the limits $\Chat\gg 1$ and $\Chat\ll 1$ allow us to write the condition for the combined foundation model to be valid as
\begin{equation}\label{eq:viscyl_faillimit}
\begin{split}
1 \ll &\frac{\Gamma(\nu)}{\alpha(\nu)(1-2\nu)}  \max\{\Chat,\Chat^{7/5}\}\\
&\equiv   \max\left\{\left[\Gamma(\nu) \frac{2R^5\rho g \sin\theta\tan\theta}{d^4 G}\right]^{2/7} ,\left[\alpha(\nu)(1-2\nu) \frac{2R^5\rho g  \sin\theta\tan\theta}{d^4G } \right]^{2/5}\right\}.
\end{split}
\end{equation}
To see this prediction in action, we turn to solving the full equilibrium equations \eqref{eq:Navier_eq} for a foundation of finite (rather than infinitesimal) depth.

\paragraph{Finite substrate depth}
If the asymptotic simplification of small aspect ratios ($\epsilon\ll 1$) is not assumed, one must solve the full two-dimensional equilibrium equations \eqref{eq:Navier_eq}  on the domain $-d \leq z\leq 0$ and $-\infty<x<\infty$. The appropriate boundary conditions are \eqref{eq:Navier_bcs} with $\bm{T}(\bm{x})\equiv \bm{0}$ and $p(\bm{x})\equiv p(x)$ given by Reynolds' equation \myeqref{eq:viscyl_eqs}{a,b}, \eqref{eq:viscyl_bcs}, and \eqref{eq:viscyl_int}. Non-dimensionalizing using the scalings presented in \eqref{eq:viscyl_dimvars} and noting that the aspect ratio can be written as
\begin{equation}\label{eq:viscyl_eps_Nnu}
\epsilon^2=\frac{d^2}{\lD^2} \equiv \frac{\alpha(\nu)(1-2\nu)}{\Gamma(\nu)\Chat},
\end{equation}
we obtain a system controlled by the Poisson ratio $\nu$, the compressibility $\Chat$, and the pliability $\phhat$, where $\Chat$ and $\phhat$ are defined in \myeqref{eq:viscyl_NphiUh}{a,b}. As in the indentation problem, the Poisson ratio now acts as an extra dimensionless parameter that cannot be scaled out of the problem.  The resulting system of partial differential equations may be solved using standard numerical techniques; details of these and the full problem are discussed in \ref{app:Numerics_example2}. 

\begin{figure}[ht!]
\centering
\includegraphics[width=0.8\textwidth]{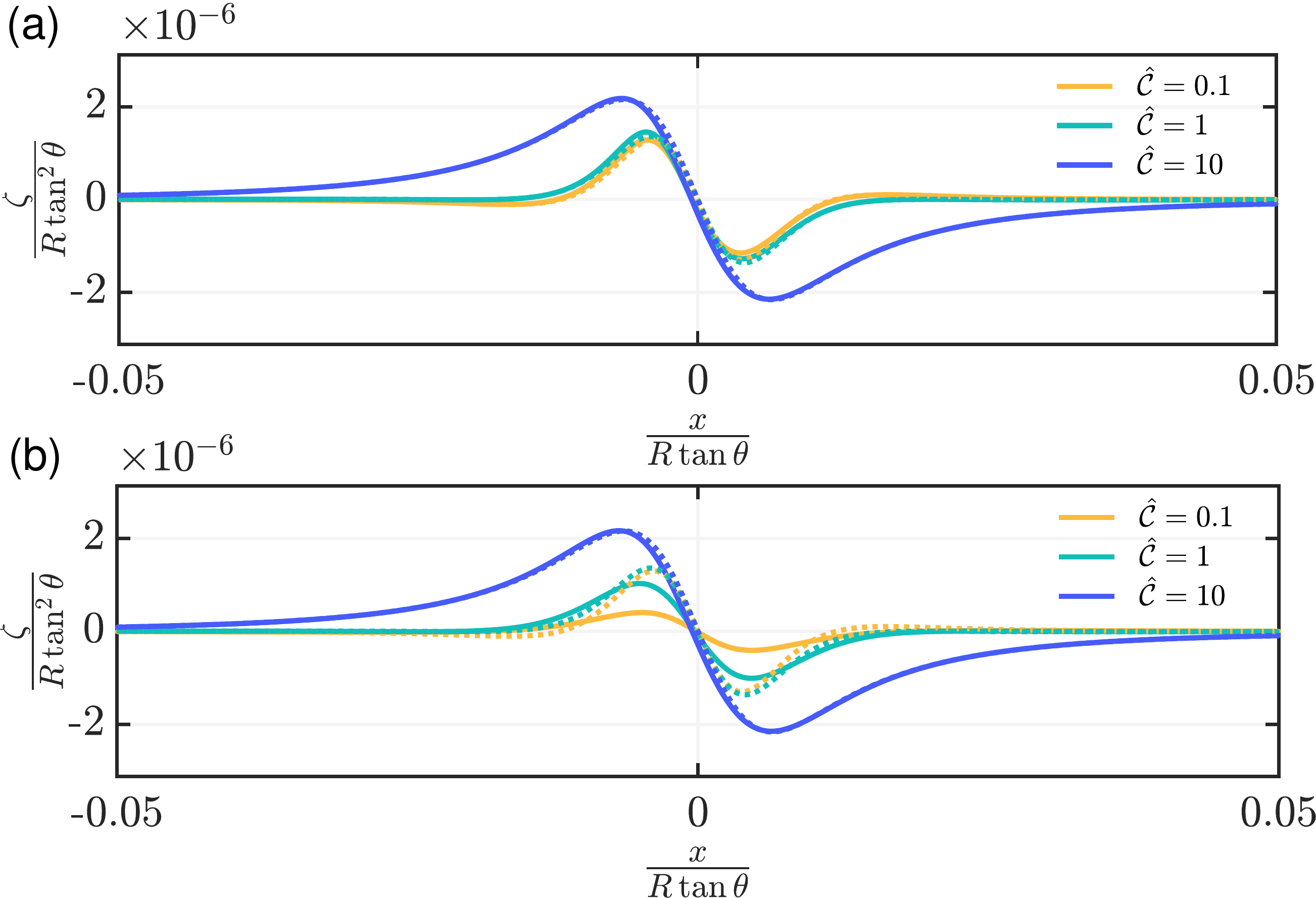}
\caption{Comparison between numerically determined surface profiles, accounting for finite foundation thickness, (solid curves) and the asymptotic predictions \eqref{eq:viscyl_asymp_profiles_Z} from the combined foundation model (dashed curves). Results are shown with $\phhat=0.01$ and different values of $\Chat$ (as described in the legend). In (a) $\nu=0.4999$, while in (b) $\nu=0.49$. Note that  disagreement between numerics and asymptotic results  is observed when $\Chat\lesssim (1-2\nu)$, as expected from \eqref{eq:viscyl_faillimit}.
\label{fig:CylIntShapes_Full}}
\end{figure}

For given values of $\nu$, $\Chat$ and $\phhat$, and hence $\epsilon$ via \eqref{eq:viscyl_eps_Nnu}, the deflection of the substrate surface, together with the equilibrium gap thickness, $\hhat$, translational speed, $\Uhat$, and rotational speed, $\ohat$, may be computed. Figure~\ref{fig:CylIntShapes_Full} shows the substrate deformation determined numerically, and compares this with the asymptotic prediction \eqref{eq:viscyl_asymp_profiles_Z}, which accurately reproduces the solution of the combined foundation model for $\phhat\ll1$ (see Fig.~\ref{fig:CylIntShapes}). Note also that even with $\nu=0.4999$, the surface profile for $\Chat=10$ in Fig.~\myref{fig:CylIntShapes_Full}{a} exhibits just the two turning points characteristic of the Winkler (rather than incompressible) response.

\begin{figure}[ht!]
\centering
\includegraphics[width=0.8\textwidth]{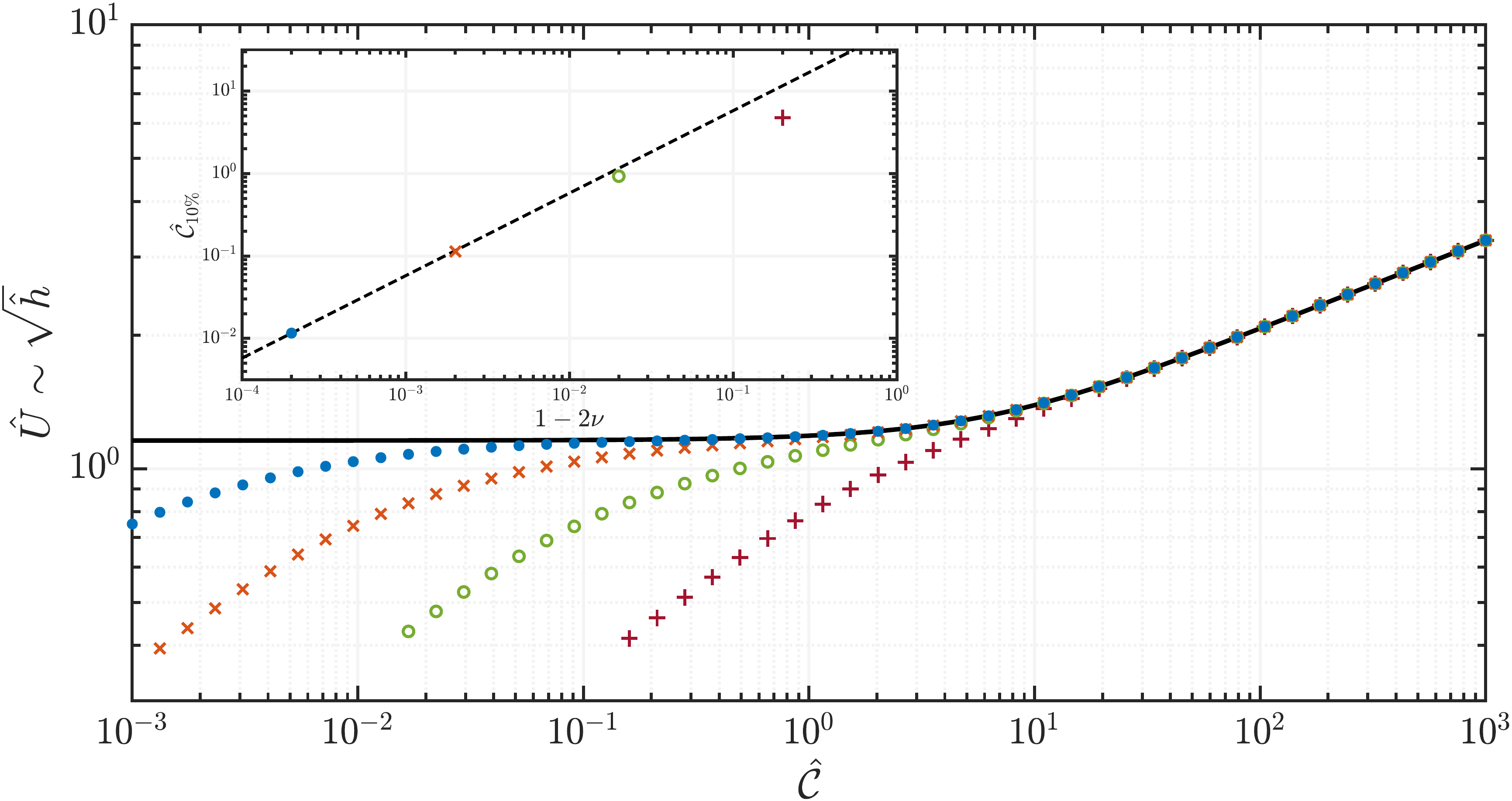}
\caption{Main figure: Comparison between the prediction of the combined foundation (solid curves) and numerical results assuming a finite-thickness foundation (points) for the sedimentation of a cylinder besides an elastomeric coating. Numerical results are obtained by solving [eqs.~\eqref{eq:applub_equilb} with \myeqref{eq:viscyl_sys_dimless}{a}] and are shown for different values of the Poisson ratio $\nu$ as follows: $\nu=0.4$ ($+$), $\nu=0.49$ ($\circ$), $\nu=0.499$ ($\times$) and $\nu=0.4999$ ($\bullet$).  We observe good agreement between the numerics and the combined foundation model provided that $\Chat\gg(1-2\nu)$. Also, note that here $\phhat=0.01\ll1$ so the $\Uhat\sim\hhat^{1/2}$ and  $\ohat \ll 1$. Inset: The $\Chat$ for which the  relative error in the use of the combined foundation model is  10\%, denoted $\Chat_{10\%}$, scales linearly with $1-2\nu$, as indicated by the dashed line; for $\C\geq\C_{10\%}$ numerical results lie within $10\%$ of the predictions of the combined foundation model. 
\label{fig:CylWithErrors}}
\end{figure}

A comparison between the numerically-determined $\Uhat$ obtained with a finite foundation thickness and that predicted by the combined foundation model is shown in Fig.~\ref{fig:CylWithErrors} for varying $\Chat$ and $\nu$, with $\phhat=0.01\ll1$ fixed.  Figures \ref{fig:CylIntShapes_Full} and \ref{fig:CylWithErrors} show   excellent agreement between the prediction of the combined foundation model and the numerical solutions of the full problem for large $\Chat$ (or, from \eqref{eq:viscyl_eps_Nnu}, small aspect ratios, $\epsilon\ll1$). However, as $\Chat$ decreases (\ie~increasing aspect ratios, $\epsilon$) we see that the numerical results  deviate from the prediction of the combined foundation model. This discrepancy becomes appreciable for $\Chat\lesssim 1-2\nu$,  consistent with the prediction of \eqref{eq:viscyl_faillimit}: results in Fig.~\ref{fig:CylWithErrors} with $\nu=0.49$ do not show the plateau in $\Uhat$ predicted by the incompressible foundation model for $\Chat\ll1$  because the small aspect ratio assumption breaks down before the incompressible limit of the combined foundation model becomes applicable. However, with $\nu=0.4999$ the expected plateau region is observed for $10^{-2}\lesssim\Chat\lesssim1$.

%====================================
Finally, we turn to discuss, briefly, the good agreement between experiments (which use approximately incompressible substrates) and theory (which is based on a model of a compressible substrate). We do not have sufficient details of the experimental parameters, particularly the Poisson ratio $\nu$, to say clearly in which regime our model would suggest these experiments lie. However, assuming that $0.45\lesssim\nu\lesssim0.49$, we find that the experiments of Saintyves \emph{et al.} \cite{Saintyves2020} appear to have $10^{-2}\lesssim\Chat\lesssim10^2$, suggesting that experiments  may span the transition between incompressible and compressible (Winkler) behaviour.  We hope that future experiments might include explicit estimates of the parameters $\Chat$ and $\nu$ allowing comparisons with the theoretical results presented here to be made.

\section{Conclusion}

In this paper we have presented an asymptotic analysis of the deformation of a thin elastic substrate in response to pressure and shear loading at its surface. This analysis led to a combined foundation model, \eqref{eq:Chandler_full}, that is able to describe the deformation of a slender foundation irrespective of how close it is to incompressible. We used this combined foundation with the assumption of zero surface-shear and zero surface-slip, \eqref{eq:Chandler_found}, to demonstrate that whether a given material behaves as an incompressible or compressible substrate is not determined solely by its Poisson ratio: even with a Poisson ratio that is arbitrarily close to $\nu=1/2$, other properties of the system may lead to behaviour that is effectively compressible, and hence may be well-described by the classical Winkler model.

Generally speaking, which of these behaviours (compressible or incompressible) is observed is determined by a compressibility parameter, which may be written
\beq
\C\propto(1-2\nu)\frac{\ell^2}{d^2},
\eeq
where $d$ is the depth of the foundation and $\ell$ is the characteristic horizontal length scale. Unfortunately, the length scale $\ell$ can only be determined by the solution of the problem at hand; in this regard, the combined foundation model can be used as a more precise alternative to scaling analysis to determine the relevant behaviour. 

We have illustrated the importance of the compressibility parameter $\C$ in two concrete examples and shown how the combined foundation model may be used to derive results valid for intermediate values of the compressibility $\C$, where scaling results are not available.  We have also shown that the combined foundation model breaks down when the aspect ratio of the foundation, $d/\ell$, becomes $\Oh(1)$, at which point the foundation behaves more akin to an elastic half-space.

A feature of the results that is, at first sight, surprising is that materials with Poisson ratio very close to $1/2$ may be well-described by the Winkler model. For example, the results in Fig.~\ref{fig:CylWithErrors} with $\nu=0.4999$ indicate that the (compressible) Winkler model is valid for extremely slender foundations provided the aspect ratio $\epsilon\lesssim 5\times10^{-3}$. 

The transition from incompressible to Winkler behaviour might also give a sensitive test of the Poisson ratio $\nu$ for materials that are close to incompressible, $\nu\approx1/2$. For example, in the first example considered here, the indentation of a stiff thin sheet coating a thin soft layer, $\C=\lW^4/\lD^4\propto (1-2\nu)(B/Gd^3)^{1/3}$. If the position of the `kink' in the dimensionless stiffness, Fig.~\ref{fig:IndentationStiffness}, can be accurately measured, and identified with $\C=\Oh(1)$ then it would be possible to obtain an accurate measure of $1-2\nu$, and hence the deviation of $\nu$ from $1/2$. For example, in Fig.~\ref{fig:IndentationStiffness}, a 10\% discrepancy between the incompressible plateau and the combined foundation stiffness  occurs at $\C\approx 0.1547$. Hence, if one is able to plot an experiment's indentation stiffness, $\KD$, for varying compressibility, $(B/Gd^3)^{1/3}$, and measure a 10\% discrepancy at $(B/Gd^3)^{1/3}=X$ say, then our results suggest that the substrate's Poisson ratio is $\nu\approx 0.5-0.4054 X$. We leave this, and other applications of the combined foundation model, to future work.

\begin{appendix}

\section{Derivation of asymptotic foundation model}\label{app:asymp_deriv}

In this Appendix, we present a detailed derivation of the asymptotic foundation model \eqref{eq:Chandler_full}, by seeking an asymptotic solution to \eqref{eq:Navier_eq} and \eqref{eq:Navier_bcs} in the limit of a small aspect ratio, $\epsilon\coloneqq d/\ell\to0$. 

We begin by noting that a geometrical argument leads to the conclusion that the stresses and strains in the plane are much smaller than their vertical counterparts \cite[Chap.~6]{Howell2008};  we therefore  assume that $[\bm{u}]\sim \epsilon [w]$ and $[\bm{T}]\sim \epsilon [p]$ to simplify our analysis, though assuming instead that $[\bm{u}]\sim[w]$ and $[\bm{T}]\sim [p]$ ultimately gives the same solution.

Based on this rescaling, we introduce the dimensionless variables
\begin{subequations}\label{eq:main_dimvars}
\begin{equation}
\bm{X} \coloneqq \frac{\bm{x}}{\ell}=\frac{\epsilon \bm{x}}{d}, \qquad 
Z \coloneqq \frac{z}{d}, \qquad \bm{U}(\bm{X},Z) \coloneqq \frac{\bm{u}(\bm{x},z)}{\epsilon[w]}, \qquad W(\bm{X},Z) \coloneqq \frac{w(\bm{x},z)}{[w]},\subtag{a--d}
\end{equation}
with dimensionless applied pressure and shear stress
\begin{equation}
P(\bm{X})\coloneqq \frac{d}{[w]}\frac{p(\bm{x})}{G} \quad \text{and} \quad \bm{\mathcal{T}}(\bm{X})\coloneqq \frac{d}{\epsilon [w]}\frac{\bm{T}(\bm{x})}{G}. \subtag{e,f}
\end{equation}
\end{subequations}

By inserting \eqref{eq:main_dimvars} into the system, \eqref{eq:Navier_eq} subject to \eqref{eq:Navier_bcs}, we obtain the dimensionless system:
\begin{subequations}\label{eq:Navier_dimless}
\begin{align}
W_{ZZ} &= -\frac{\epsilon^2}{2(1-\nu)}\left(\nabh\cdot\bm{U}_Z+(1-2\nu)\nabh^2W \right),\\
\bm{U}_{ZZ} +\frac{\nabh W_Z}{1-2\nu}&= -\frac{\epsilon^2}{1-2\nu}\left(\nabh\left(\nabh\cdot\bm{U}\right)+(1-2\nu)\nabh^2\bm{U}\right),
\end{align}
\end{subequations}
in $-1<Z<0$, subject to the boundary conditions:
\begin{subequations}
\begin{align}
W=0 \quad \text{and}  \quad \bm{U}=\bm{0} &&&\quad \text{on $Z=-1$,}\subtag{a,b}\\
W_Z+\frac{\nu\epsilon^2}{1-\nu}\nabh \cdot\bm{U}=-\frac{1-2\nu}{2(1-\nu)}P(\bm{X}) \quad \text{and}  \quad
\bm{U}_Z+\nabh W =\bm{\mathcal{T}}(\bm{X}) &&&\quad \text{on $Z=0$.}\subtag{c,d}
\end{align}
\end{subequations}
Assuming $(1-2\nu) = \Oh(1)$, we seek asymptotic solutions of the system by inserting the expansions,
\begin{subequations}\label{eq:asymp_exp}
\begin{align}
W(\bm{X},Z)&= W^{(0)}(\bm{X},Z) +\epsilon^2 W^{(1)}(\bm{X},Z)+\Oh(\epsilon^4),\\
\bm{U}(\bm{X},Z)&= \bm{U}^{(0)}(\bm{X},Z) +\epsilon^2\bm{U}^{(1)}(\bm{X},Z)+\Oh(\epsilon^4),
\end{align}
\end{subequations}
 and solving order-by-order as $\epsilon\to0$.

\paragraph{At leading-order, $\Oh(\epsilon^0)$.} The displacements $W^{(0)}(\bm{X},Z)$ and $\bm{U}^{(0)}(\bm{X},Z)$ satisfy,
\begin{subequations}
\begin{equation}
W^{(0)}_{ZZ} = 0\quad \text{and} \quad \bm{U}^{(0)}_{ZZ} =-\frac{1}{1-2\nu}\nabh W^{(0)}_Z,\subtag{a,b}
\end{equation}
\end{subequations}
in $-1<Z<0$, subject to the boundary conditions
\begin{subequations}
\begin{align}
\bm{U}^{(0)}(\bm{X},-1)=\bm{0}, &\qquad W^{(0)}(\bm{X},-1)=0,\subtag{a,b}\\
W^{(0)}_Z(\bm{X},0)=-\frac{1-2\nu}{2(1-\nu)}P(\bm{X}), &\qquad
\bm{U}^{(0)}_Z(\bm{X},0)+\nabh W^{(0)}(\bm{X},0) =\bm{\mathcal{T}}(\bm{X}).\subtag{c,d}
\end{align}
\end{subequations}
Integrating with respect to $Z$ leads to the solutions:
\begin{subequations}\label{eq:U0W0_sol}
\begin{align}
\bm{U}^{(0)}(\bm{X},Z)&= \left(1+Z\right)\bm{\mathcal{T}}(\bm{X})-\frac{4\nu-1}{4(1-\nu)}\left(1+Z\right)\left[1-\frac{Z}{4\nu-1}\right]\nabh P(\bm{X}),\\
W^{(0)}(\bm{X},Z)&= -\frac{1-2\nu}{2(1-\nu)}\left(1+Z\right)P(\bm{X}).
\end{align}
\end{subequations}
These describe the leading-order displacements provided $(1-2\nu)=\Oh(1)$, and hence reproduce the results of the Winkler limit given by \cite{Skotheim2004}. However,  as $\nu\to 1/2$ (\ie~in the incompressible limit)  $W^{(0)}(\bm{X},Z)\to 0$ and so higher-order terms dominate; in particular, $\bm{U}\sim \bm{U}^{(0)}$ and $W\sim \epsilon^2W^{(1)}$ for $(1-2\nu)\ll\epsilon^2$ where we  determine $W^{(1)}$ next.

\paragraph{At next-order, $\Oh(\epsilon^2)$.} The displacement $W^{(1)}(\bm{X},Z)$ satisfies,
\begin{equation}
W^{(1)}_{ZZ}=-\frac{1}{2(1-\nu)}\left(\nabh\cdot\bm{U}^{(0)}_Z+(1-2\nu)\nabh^2W^{(0)} \right),
\end{equation}
in $-1<Z<0$, subject to the boundary conditions
\begin{subequations}
\begin{equation}
W^{(1)}(\bm{X},-1)=0\quad \text{and} \quad W^{(1)}_Z(\bm{X},0)=-\frac{\nu}{1-\nu}\nabh \cdot\bm{U}^{(0)}(\bm{X},0).\subtag{a,b}
\end{equation}
\end{subequations}
Substituting the known solutions for $\bm{U}^{(0)}$ and $W^{(0)}$, and integrating with respect to $Z$, leads to:
\begin{equation}\label{eq:W1_sol}
\begin{split}
W^{(1)}(\bm{X},Z)= &-\frac{4\nu-1}{4(1-\nu)}\left(1+Z\right)\left[1+\frac{Z}{4\nu-1}\right]\nabh \cdot\bm{\mathcal{T}}(\bm{X})\\
&+\frac{\nu(4\nu-1)}{6(1-\nu)^2}\left(1+Z\right)\left[1+\frac{Z}{2}-\frac{1-\nu}{4\nu-1}Z^2\right]\nabh^2 P(\bm{X}).
\end{split}
\end{equation}
\paragraph{Overall.} Combining \eqref{eq:U0W0_sol} with \eqref{eq:W1_sol}, and changing back to dimensional variables, we find expressions for the  substrate displacements accurate to $\Oh(\epsilon^2d\bm{T}/G,\,\epsilon^2 d^2\nabh p/G)$ for $\bm{u}$ and  $\Oh(\epsilon^2d^2\nabh\cdot\bm{T},\,\epsilon^2d^3\nabh^2 p/G)$ for $w$. Substituting  $z=0$ yields the surface displacement expressions given in the main text,  \eqref{eq:Chandler_full}.

%\numberwithin{equation}{section} % Needed to sort out the strange Appendix eqution numbering

\section{Numerical solutions of the finite depth foundation}\label{app:Numerics}

In this Appendix, we present further details of the numerical simulations performed to model an elastic foundation with finite thickness. We present the governing equations for the examples of \S\ref{sec:FirstExample} and \S\ref{sec:SecondExample} and discuss the numerical solutions of these systems.

\subsection{First example: A stiff coating}\label{app:Numerics_example1}

In \S\ref{sec:FirstExample} we considered the point-indentation of a soft substrate with a stiff coating. Using the scalings \eqref{eq:stiffcoat_dimvars} and \eqref{eq:main_dimvars}, we introduce  dimensionless coordinates, displacements, and pressure:
\begin{subequations}
\begin{equation}
\rho \coloneqq \frac{r}{\lD},\quad\enskip
Z \coloneqq \frac{z}{d}, \quad\enskip
 U(\rho,Z) \coloneqq \frac{u(r,z)}{\epsilon\delta}, \quad\enskip
  W(\rho,Z) \coloneqq \frac{w(r,z)}{\delta},\quad\enskip
Q(\rho) \coloneqq \frac{\lD^4q(r)}{B\delta}, \subtag{a--e}
\end{equation}
\end{subequations}
where $\lD$ is the horizontal length-scale for an incompressible foundation,  \myeqref{eq:stiffcoat_LDLW}{b}, and $\epsilon=d/\lD$ is the aspect ratio, \eqref{eq:stiffcoat_eps_Nnu}.  With this non-dimensionalization, the  displacements $U(\rho,Z)$ and $W(\rho,z)$ satisfy:
\begin{subequations}\label{eq:appStiff_equil}
\begin{align}
W_{ZZ} &= -\frac{\epsilon^2}{2(1-\nu)}\left(\frac{1}{\rho}\frac{\partial(\rho U_Z)}{\partial\rho}+(1-2\nu)\frac{1}{\rho}\frac{\partial(\rho W_\rho)}{\partial\rho} \right),\\
U_{ZZ} +\frac{ W_{\rho Z}}{1-2\nu}&= -\frac{2(1-\nu)}{1-2\nu}\epsilon^2\frac{\partial}{\partial\rho}\left(\frac{1}{\rho}\frac{\partial(\rho U)}{\partial\rho}\right),
\end{align}
\end{subequations}
in $-1<Z<0$ and $0<\rho<\infty$, subject to the boundary conditions:
\begin{subequations}\label{eq:appStiff_bcs}
\begin{align}
W,\, U\to 0 &&&\qquad \text{as $\rho\to\infty$,}\subtag{a,b}\\
W_\rho=0 \quad \text{and}  \quad U=0 &&&\qquad \text{on $\rho=0$,}\subtag{c,d}\\
W=0 \quad \text{and}  \quad U=0 &&&\qquad \text{on $Z=-1$,}\subtag{e,f}\\
W_Z+\frac{\nu\epsilon^2}{1-\nu}\frac{1}{\rho}\frac{\partial(\rho U)}{\partial\rho}= -\C Q(\rho) \quad \text{and}  \quad
U=0 &&&\qquad \text{on $Z=0$,}\subtag{g,h}\\
W=-1&&&\qquad \text{at $Z=\rho=0$.} \subtag{i}
\end{align}
\end{subequations}
where $Q(\rho)$ is given by the Kirchoff-Love plate equation: 
\begin{equation}\label{eq:appStiff_plate}
Q(\rho) = \nabh^4 W(\rho,0) +\KD\frac{\dirac(\rho)}{\rho}.
\end{equation}
For given values of $\nu$ and $\C$,  \eqref{eq:appStiff_equil}--\eqref{eq:appStiff_plate} describe a system for the substrate displacements $U(\rho,Z)$ and $W(\rho,Z)$ with the problem for $W(\rho,0)$ over-determined determining the stiffness $\KD$.

The system \eqref{eq:appStiff_equil}--\eqref{eq:appStiff_plate} can be solved by  standard numerical techniques;  we use the method of finite differences after first eliminating $U(\rho,Z)$ by cross-differentiating, to obtain the biharmonic formulation of the elastostatic equations. This gives a fourth-order PDE for $W(\rho,z)$, instead of a second-order PDE system for $W(\rho,z)$ and $U(\rho,Z)$.

\subsection{Second example: A lubrication problem}\label{app:Numerics_example2}
In \S\ref{sec:SecondExample} we considered the equilibrium sliding of an infinite cylinder above an inclined soft substrate within a viscous liquid. Using the scalings  \eqref{eq:viscyl_dimvars} and \eqref{eq:main_dimvars}, we non-dimensionalize by letting
\begin{subequations}
\begin{align}
X &\coloneqq  \frac{x}{\lD}, & U(X,Z) &\coloneqq  \frac{2R^2\tan\theta}{\lD^2 d}u(x,z),  & \Hcal(X) &\coloneqq \frac{2R  H(x)}{\lD^2},  \subtag{a--c}\\
Z &\coloneqq \frac{z}{d}, &  W(X,Z) &\coloneqq \frac{2R^2\tan\theta}{\lD^3}w(x,z),& P(X) &\coloneqq \frac{\lD^2 p(x)}{R^3 \rho g \sin\theta},  \subtag{d--f}
\end{align}
\end{subequations}
where $\lD$ is the horizontal length-scale  associated with an incompressible response, defined in \myeqref{eq:viscyl_LDLW}{b}.  The dimensionless foundation displacements $U(X,Z)$ and $W(X,Z)$ then satisfy:
\begin{subequations}\label{eq:applub_equilb}\begin{align}
W_{ZZ} &= -\frac{\epsilon^2}{2(1-\nu)}\left[U_{XZ}+(1-2\nu)W_{XX} \right],\\
U_{ZZ} +\frac{W_{XZ}}{1-2\nu}&= -\frac{2(1-\nu)}{1-2\nu}\epsilon^2U_{XX},
\end{align}\end{subequations}
in $-1<Z<0$ and  $-\infty<X<\infty$, subject to the boundary conditions:
\begin{subequations}\label{eq:applub_bcs}
\begin{align}
W,\, U\to 0 &&&\qquad \text{as $X\to \pm \infty$,}\subtag{a,b}\\
%W=0 \quad \text{and}  \quad U=0 &&&\qquad \text{on $Z=-1$,}\\
W_Z+\frac{\nu\epsilon^2}{1-\nu}U_X=-\Chat P(X) \quad \text{and}  \quad
U_Z+ W_X = 0&&&\qquad \text{on $Z=0$,}\subtag{c,d}
\end{align}
\end{subequations}
and \myeqref{eq:appStiff_bcs}{e,f} where $\epsilon=d/\lD$ is the aspect ratio, \eqref{eq:viscyl_eps_Nnu}, and $P(X)$ satisfies Reynolds' equation \myeqref{eq:viscyl_sys_dimless}{a} with $\Hcal(X) = \hhat+X^2 -\phhat W(X,0)$; this is to be solved
subject to $P(X) \to 0$ as $X\to \pm \infty$ and the integral constraints \myeqref{eq:viscyl_sys_dimless}{d--f}.

Overall, given values for the dimensionless parameters $\nu$, $\Chat$, and $\phhat$,  \eqref{eq:applub_equilb}--\eqref{eq:applub_bcs} with \myeqref{eq:viscyl_sys_dimless}{a,b} and \myeqref{eq:viscyl_sys_dimless}{d--f} describe a system for the substrate displacements $U(X,Z)$ and $W(X,Z)$ and the applied pressure $P(X)$; the gap thickness, $\hhat$, translational speed, $\Uhat$, and rotational speed, $\ohat$, are determined as part of the solution since the problem for the pressure field $P$ is over-determined. This system can be solved using standard numerical techniques; we use the method of finite differences and numerical quadrature with the nonlinear system solver \texttt{fsolve} in \texttt{\textsc{MATLAB}}.

\end{appendix}

\enlargethispage{20pt}

%\ethics{Insert ethics text here.}

%\dataccess{Data associated with this paper will be uploaded to the Oxford Research Archive.}

%\aucontribute{TGJC carried out the numerical and asymptotic analysis. DV conceived of and designed the study. Both authors drafted and approved the manuscript.}

%\competing{We declare that we have no competing interests.}

\section*{Acknowledgments}
The research leading to these results has received funding from the European Research Council under the European Union's Horizon 2020 Programme/ERC Grant No.~637334 (D.V.), the Leverhulme Trust (D.V.) and the EPSRC Grant No.~EP/M508111/1 (T.G.J.C.).  We are grateful to Finn Box, Chris MacMinn and Satyajit Pramanik  for discussions about this work.
%\end{acknowledgements}

%\ack{}

%\disclaimer{Insert disclaimer text here.}

%%%%%%%%%% Insert bibliography here %%%%%%%%%%%%%%

\vskip2pc

%% Use bibtex to generate bibliography, then copy paste to here if necessary

%\bibliographystyle{RS} %%%% .BST file
%
%\bibliography{CompositeFoundation} %%%%% .Bib file

\end{document}